\documentclass[pdflatex,sn-mathphys-num]{sn-jnl}% Math and Physical Sciences Numbered Reference Style
\usepackage{graphicx}%
\usepackage{multirow}%
\usepackage{amsmath,amssymb,amsfonts}%
\usepackage{amsthm}%
\usepackage{mathrsfs}%
\usepackage[title]{appendix}%
\usepackage[table]{xcolor}%
\usepackage{tcolorbox}
\usepackage{textcomp}%
\usepackage{manyfoot}%
\usepackage{booktabs}%
\usepackage{algorithm}%
\usepackage{algorithmicx}%
\usepackage{algpseudocode}%
\usepackage{listings}%
\usepackage{makecell}

\theoremstyle{thmstyleone}%
\theoremstyle{thmstyletwo}%

\theoremstyle{thmstylethree}%
\newcommand{\grayrow}{\rowcolor[gray]{.9}}
\colorlet{darkgreen}{green!65!black}
\begin{document}

\title[Article Title]{Camera Measurement of Blood Oxygen Saturation}

\author[1]{\fnm{Jiankai} \sur{Tang}}\email{tjk24@mails.tsinghua.edu.cn}
\author[2]{\fnm{Xin} \sur{Liu}}
\author*[2]{\fnm{Daniel} \sur{McDuff}}\email{dmcduff@uw.edu}
\author[1]{\fnm{Zhang} \sur{Jiang}}
\author[1]{\fnm{Hongming} \sur{Hu}}
\author[1]{\fnm{Luxi} \sur{Zhou}}
\author[3]{\fnm{Nodoka} \sur{Nagao}}
\author[3]{\fnm{Haruta} \sur{Suzuki}}
\author[3]{\fnm{Yuki} \sur{Nagahama}}
\author[1]{\fnm{Wei} \sur{Li}}
\author[1]{\fnm{Linhong} \sur{Ji}}
\author[1]{\fnm{Yuanchun} \sur{Shi}}
\author*[3]{\fnm{Izumi} \sur{Nishidate}}\email{inishi@cc.tuat.ac.jp}
\author*[1]{\fnm{Yuntao} \sur{Wang}}\email{yuntaowang@tsinghua.edu.cn}

\affil[1]{Tsinghua University}
\affil[2]{University of Washington}
\affil[3]{Tokyo University of Agriculture and Technology}


\abstract{Blood oxygen saturation (SpO$_2$) is a critical vital sign routinely monitored in medical settings. Traditional measurement requires dedicated contact sensing equipment. This study introduces a deep learning framework for contactless SpO$_2$ measurement using an off-the-shelf camera, addressing the challenges of real-world variability in lighting and skin tone. We conducted two large-scale studies with diverse participants, comparing our method’s performance against traditional signal processing approaches in both \emph{intra-} and \emph{inter-} dataset settings. Our approach achieved consistent accuracy across varied demographic groups, demonstrating the feasibility of camera-based SpO$_2$ monitoring as a scalable, non-invasive tool for remote health assessment.
}




\maketitle

\section{Introduction}
\label{sec:introduction}

Assessment of physiological vitals is a necessary part of health care. The most common measures are body temperature, heart rate, blood pressure, and blood oxygen saturation. Although these signals are typically measured using different medical devices, the field of ubiquitous health has endeavored to find methods of doing so using everyday devices. Cameras are one of the most versatile and available sensors on everyday electronics and can measure spatial, temporal signals across the visual section of the electromagnetic spectrum. Camera measurement of vital signs leverages imaging devices to compute physiological changes by analyzing images of the human body~\cite{mcduff2023camera}.  

Photoplethysmography and pulse rate measurement have received the majority of attention~\cite{blazek2000near,verkruysse2008remote,takano2007heart,wang2017algorithmic,chen2018deepphys,tang2024camera,castellanoontiverosMachineLearningbasedApproach2024}. Deep learning methods provide the state of the art for this task~\cite{chen2018deepphys,mcduff2018deep,yu2019remote,liu2020multi,yu2022physformer} thanks to the ability of these highly parameterized networks to capture subtle changes while ignoring different sources of noise~\cite{chen2020deepmag,nowara2021benefit,tang2023mmpd}. As with any task machine larning models, based on training data, are a function of the data used to create them. In the case of remote physiological measurement models are typically trained using labels obtained from contact ``gold standard'' sensors.  

\begin{figure}[t!]
    \centering
    \includegraphics[width=0.9\textwidth]{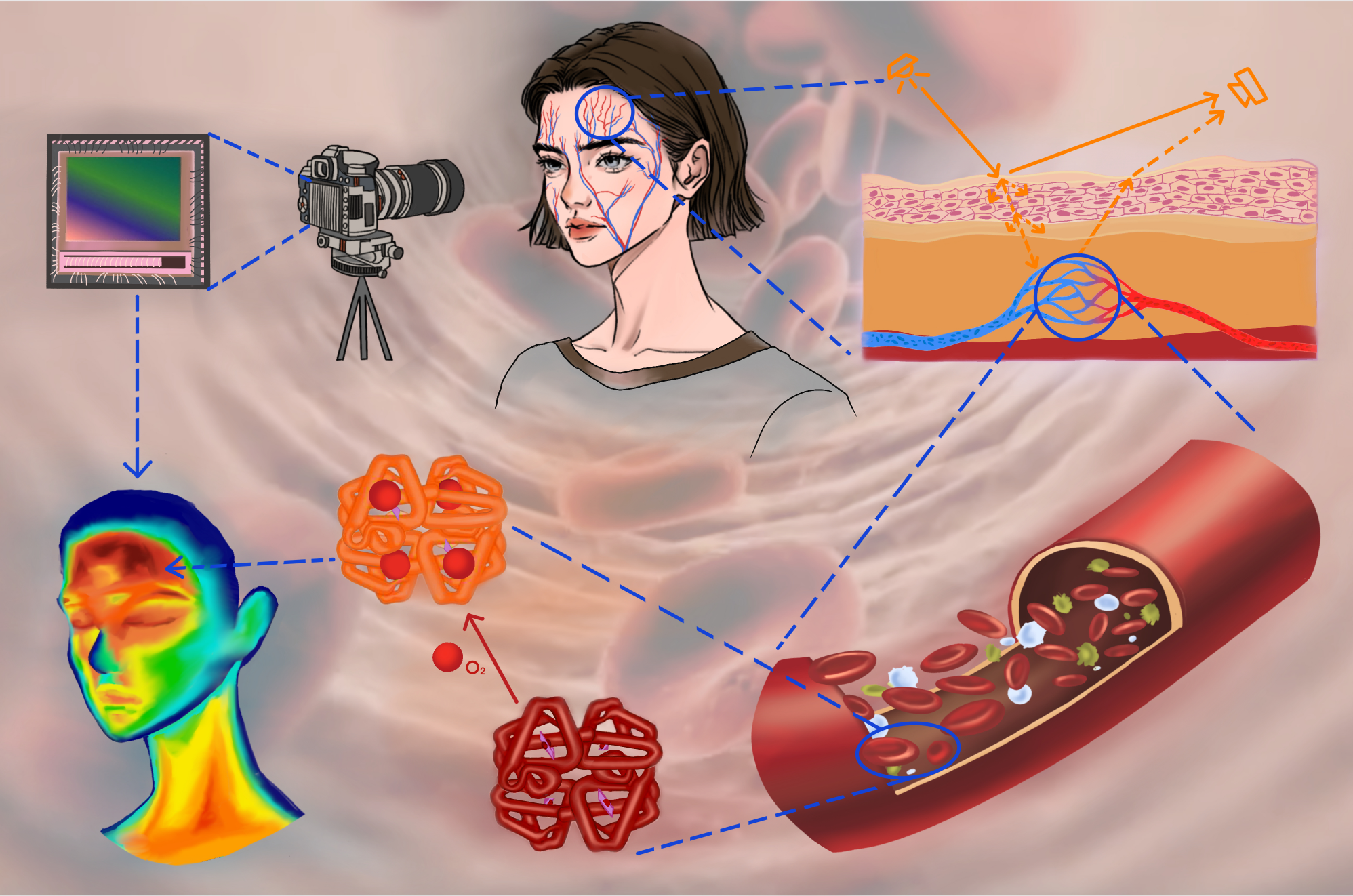}
    \vspace{0.1cm}
    \caption{\textbf{Principle of Camera-based SpO$_2$ Measurement.} The camera captures the skin tissue, which is illuminated by a light source. Beneath the skin tissue, blood vessels contain oxygenated and deoxygenated hemoglobin. The RGB values in each pixel can be transformed into the concentrations of oxygenated hemoglobin (HbO) and the deoxygenated hemoglobin (HbR) using a skin tissue model~\cite{nishidate2022rgb}. Then the tissue oxygen saturation (StO$_2$) and the percutaneous arterial oxygen saturation (SpO$_2$) can be calculated from the HbO and HbR concentrations.}
    \label{fig:Teaser}
\end{figure}

Peripheral blood oxygen saturation is an important and consequential vital sign that reflects respiratory functioning. Blood oxygen is regulated with precision across the body due to the risk posed to critical organs such as the heart and brain if it drops~\cite{hafen2018oxygen}. 
The COVID-19 pandemic led to increased interest in scalable, non-invasive methods for measuring SpO$_2$. 
Camera measurement of oxygen levels is challenging due to the fact that color sensitivity profiles for a given camera may not be known, even if they are the red, green and blue bands have wide absorption profiles meaning that they cannot be used to measure specific frequencies precisely. Furthermore, ambient lighting (intensity, color) in the environment is rarely known or precisely controllable. As such algorithms either need to be carefully calibrated using known reference colors or perform that inference automatically. Fig.~\ref{fig:Teaser} shows the principle of camera-based SpO$_2$ measurement.

Computer vision approaches for measuring blood oxygen levels have been designed~\cite{shao2015noncontact,guazzi2015non,mathew2023remote,ding2018measuring,hoffman2022smartphone,nishidate2022rgb}. However, these systems often rely on custom hardware~\cite{shao2015noncontact}, are evaluated on a small number of subjects (e.g., N=5~\cite{guazzi2015non}, N=10~\cite{liu2024summit}, N=14~\cite{mathew2023remote}) or require the body to be in contact with the camera (i.e., are not remote)~\cite{ding2018measuring,hoffman2022smartphone}. There are a few notable studies with larger populations and creative study designs. Wu et al.~\cite{wu2023peripheral} performed analysis of 60 subjects at altitude. However, their results are only based on a leave-one-subject-out (LOSO) design. Van Gastel et al.~\cite{van2022contactless} collected two datasets to test performance in realistic spot check scenarios. They determined calibration coefficients on one dataset that they showed could be made acceptable for realistic screening applications on the second population.

The approaches fall into two categories: signal processing methods based on principled assumptions of the underlying signals~\cite{nishidate2022rgb} and deep learning methods in which the mapping between pixels and SpO$_2$ are learned from a training set. The former explicitly contains parameters for calibration; however, can a machine learning approach do without such calibration?  The evidence of generalization is limited, and many methods struggle to produce convincing results~\cite{mathew2023remote} or are evaluated on data sets with limited variation in oxygenation~\cite{cheng2024contactless}.

As cameras vary in properties and different environments some form of calibration of the camera to the ambient illumination conditions is a necessary step. In this paper, we investigate the feasibility of contactless SpO$_2$ measurement using video across diverse settings and populations, addressing key factors such as lighting variability, skin tone, and the need for calibration to ensure reliable and accurate readings. Specifically we focus on whether supervised training can lead to good results in a cross-domain application and how much, if any, calibration is needed to achieve acceptable performance. We summarize our contributions as follows:

\begin{enumerate} 
    \item Conduct two studies with diverse participants to validate the feasibility of contactless blood oxygen saturation measurement via consumer-grade cameras under varied environmental conditions. 
    \item Develop an end-to-end deep learning framework that combines video data with color calibration for accurate and contactless SpO$_2$ estimation. 
    \item Demonstrate the model’s performance through rigorous \emph{intra-} and \emph{inter-} dataset evaluations, achieving substantial improvements over traditional signal processing approaches in diverse settings. 
    \item Perform subgroup analyses to investigate the effects of demographic factors, including skin tone, age, gender, and COVID-19 history, on measurement accuracy, establishing baseline insights for equitable performance across populations. 
\end{enumerate}

% Skin type is known to impact the performance of optical measurement of physiological signals~\cite{nowara2020meta}. However, many studies fail to recruit a diverse population in order to assess the performance of with respect to the skin tone. In addition to recruiting a larger number of subjects than in previous work, we created a more diverse ...

\section{Results}
\label{sec:results}

\begin{figure}[t!]
    \centering
    \includegraphics[width=1\textwidth]{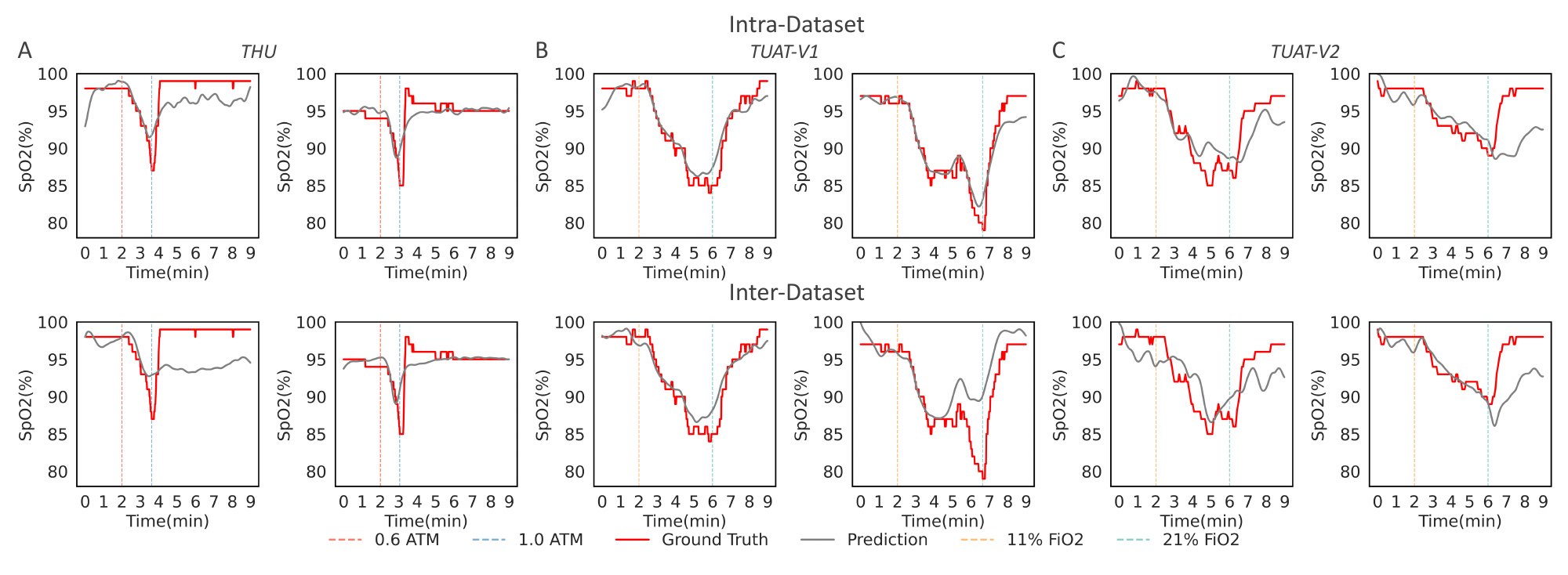} %\\
    \vspace{0.2cm}
    \caption{\textbf{Examples of Predicted SpO2 and Ground Truth SpO2 across three datasets and two evaluation setups.} A) Intra-dataset evaluation and Inter-dataset evaluation on the THU dataset. B) Intra-dataset evaluation and Inter-dataset evaluation on the TUAT V1 dataset. C) Intra-dataset evaluation and Inter-dataset evaluation on the TUAT V2 dataset.}
    \label{fig:images}
  \end{figure}

\subsection{Overall Results}
To validate the feasibility of remote blood oxygen measurement, we conducted a series of intra-dataset, cross-dataset, and ablation experiments on our method and baseline approaches. Table \ref{Table: Dataset Description} provides details of the datasets included in both the intra-dataset and cross-dataset experiments. For the ablation study, we selected the TUAT-v1 dataset to examine the effects of the number of calibration frames, alpha selection, calibration frame sampling strategy, and training labels. For the ablation study we focused on the use of a color checker, we used the THU dataset, in which the color checker was recorded in each frame alongside the subject’s face. Additionally, we analyzed the influence of potential confounders — such as skin tone, age, gender, and COVID-19 history — on the performance of our method.

% Table 1: Intra-dataset
% \usepackage{colortbl}
\textbf{Intra-dataset Experiments.}
In order to evaluate intra-dataset performance we conducted three \emph{leave-one-subject-out} (LOSO) cross-validation experiments, one on each of the THU, TUAT V1 and TUAT V2 datasets. We compared the root mean squared error (RMSE) and mean average percentage error (MAPE) results against the baseline (see Table~\ref{Table: Intra Dataset}). Our deep learning method outperforms the baseline in most settings, showing 42.4\% improvement on the TUAT V1 dataset when calibrated on 270 frames.

\begin{table}[htp]
\centering
\caption{\textbf{Intra-dataset Performance.} Leave-one-subject-out cross-validation (LOSOCV) results on the THU, TUAT V1, and TUAT V2 datasets.}
\label{Table: Intra Dataset}
\begin{tabular}{llllll} 
\toprule[1.5pt]
Calibration Frames && \multicolumn{2}{c}{50\% Frames (270)} & \multicolumn{2}{c}{1\% Frames (5)} \\ 

Dataset & Method & RMSE (\%) & MAPE (\%) & RMSE (\%) & MAPE (\%) \\ 
\hline \hline

\grayrow
THU & Baseline & 1.71 $\pm$ 0.10 & 1.46 $\pm$ 0.09 & 2.27 $\pm$ 0.11 & 1.87 $\pm$ 0.09 \\ 
 & Ours & 1.69 $\pm$ 0.11 & 1.40 $\pm$ 0.09 & 2.33 $\pm$ 0.12 & 1.94 $\pm$ 0.10 \\

\grayrow
& \textsc{Delta} &  \textcolor{darkgreen}{\texttt{-}\textbf{1.2\%}} & \textcolor{darkgreen}{\texttt{-}\textbf{4.1\%}} & \textcolor{red}{\texttt{+}\textbf{2.6\%}} & \textcolor{red}{\texttt{+}\textbf{3.7\%}} \\ \hline

TUAT V1 & Baseline & 5.62 $\pm$ 0.27 & 5.31 $\pm$ 0.27 & 7.27 $\pm$ 0.37 & 6.17 $\pm$ 0.34 \\ 
\grayrow
 & Ours & 3.24 $\pm$ 0.25 & 2.77 $\pm$ 0.23 & 5.85 $\pm$ 0.34 & 5.16 $\pm$ 0.33 \\ 

& \textsc{Delta} &  \textcolor{darkgreen}{\texttt{-}\textbf{42.4\%}} & \textcolor{darkgreen}{\texttt{-}\textbf{47.8\%}}  & \textcolor{darkgreen}{\texttt{-}\textbf{19.5\%}} & \textcolor{darkgreen}{\texttt{-}\textbf{16.4\%}}  \\ \hline

\grayrow
TUAT V2 & Baseline & 5.52 $\pm$ 0.53 & 5.06 $\pm$ 0.50 & 7.16 $\pm$ 0.68 & 5.91 $\pm$ 0.63 \\
 & Ours & 4.47 $\pm$ 1.09  & 4.05 $\pm$ 1.00 & 5.88 $\pm$ 1.42  & 4.75 $\pm$ 1.19 \\

\grayrow
 & \textsc{Delta} &  \textcolor{darkgreen}{\texttt{-}\textbf{19.0\%}} & \textcolor{darkgreen}{\texttt{-}\textbf{20.0\%}} & \textcolor{darkgreen}{\texttt{-}\textbf{17.8\%}} & \textcolor{darkgreen}{\texttt{-}\textbf{19.6\%}} \\ 

\bottomrule[1.5pt]
\end{tabular}
\arrayrulecolor{black}
\end{table}

\textbf{Inter-dataset Experiment.}
Next, we examine the inter-dataset performance and contrast this with the intra-dataset performance. Our method achieves a 5.2\%-26.9\% improvement in performance on the TUAT series but failed to generalize on the THU dataset. One reason for this is the differences in recording environment and devices used between THU and TUAT.

\begin{table}[htp]
\centering
\caption{\textbf{Inter-dataset Performance.} Inter-dataset results on the THU, TUAT V1, and TUAT V2 datasets when training on all videos from the other two datasets.}
\label{Table: Inter Dataset}
\begin{tabular}{llllll} 
\toprule[1.5pt]
Calibration Frames && \multicolumn{2}{c}{50\% Frames (270)} & \multicolumn{2}{c}{1\% Frames (5)} \\ 

Dataset & Method & RMSE (\%) & MAPE (\%) & RMSE (\%) & MAPE (\%) \\ 
\hline \hline

\grayrow
THU & Baseline & 1.71 $\pm$ 0.10 & 1.46 $\pm$ 0.09 & 2.27 $\pm$ 0.11 & 1.87 $\pm$ 0.09 \\ 
 & Ours & 1.74 $\pm$ 0.11 & 1.44 $\pm$ 0.09 & 2.47 $\pm$ 0.15 & 2.08 $\pm$ 0.14 \\

\grayrow
& \textsc{Delta} &  \textcolor{red}{\texttt{+}\textbf{1.8\%}} & \textcolor{darkgreen}{\texttt{-}\textbf{1.4\%}} & \textcolor{red}{\texttt{+}\textbf{8.8\%}} & \textcolor{red}{\texttt{+}\textbf{11.2\%}} \\ \hline

TUAT V1 & Baseline & 5.62 $\pm$ 0.27 & 5.31 $\pm$ 0.27 & 7.27 $\pm$ 0.37 & 6.17 $\pm$ 0.34 \\ 
\grayrow
 & Ours & 4.11 $\pm$ 0.36 & 3.62 $\pm$ 0.32 & 6.89 $\pm$ 0.38 & 5.83 $\pm$ 0.34 \\ 

& \textsc{Delta} &  \textcolor{darkgreen}{\texttt{-}\textbf{26.9\%}} & \textcolor{darkgreen}{\texttt{-}\textbf{31.8\%}}  & \textcolor{darkgreen}{\texttt{-}\textbf{5.2\%}} & \textcolor{darkgreen}{\texttt{-}\textbf{5.5\%}}  \\ \hline

\grayrow
TUAT V2 & Baseline & 5.52 $\pm$ 0.53 & 5.06 $\pm$ 0.50 & 7.16 $\pm$ 0.68 & 5.91 $\pm$ 0.63 \\
 & Ours & 4.49 $\pm$ 0.45 & 3.53 $\pm$ 0.34 & 6.77 $\pm$ 0.92 & 5.91 $\pm$ 0.95 \\

\grayrow
 & \textsc{Delta} &  \textcolor{darkgreen}{\texttt{-}\textbf{18.7\%}} & \textcolor{darkgreen}{\texttt{-}\textbf{30.2\%}} & \textcolor{darkgreen}{\texttt{-}\textbf{5.4\%}} & \textcolor{darkgreen}{\texttt{0.0\%}} \\ 

\bottomrule[1.5pt]
\end{tabular}
\arrayrulecolor{black}
\end{table}

\subsection{Ablation Experiment}

Next we examine the optimal parameter settings with a focus on calibration frame selection, alpha tuning, the use of accurate training labels, and the implementation of a color checking mechanism.

\textbf{Optimal Frame Selection.}
From the data presented in Table \ref{Table: Calibration Frames}, using 50\% of frames (270 frames) for calibration resulted in the lowest Mean Absolute Error (MAE) of 2.53 ($\pm$0.20) and RMSE of 3.24 ($\pm$0.25). This is a decrease in MAE of 38.9\% and RMSE of 34.35\% compared to using just 1\% of frames (5 frames) for calibration. It demonstrates that choosing a greater number of frames enhances the model's accuracy but would lead to a longer calibration time.

\begin{table}[htp]
\centering
\caption{\textbf{Ablation Study on Frame Selection with TUAT V1} Performance metrics when calibrated by varying percentages of frames on the TUAT V1 dataset with predicted SpO2 and color check.}
\label{Table: Calibration Frames}
\begin{tabular}{llllll} 
\toprule[1.5pt]
% Dataset && \multicolumn{4}{c}{TUAT V1} \\ 
Calibration & Method & MAE (\%) & RMSE (\%) & MAPE (\%) & Pearson \\ 
\hline \hline

\grayrow
First 50\% (270 frames) & Baseline & 4.68 $\pm$ 0.22 & 5.62 $\pm$ 0.27 & 5.31 $\pm$ 0.27 & - \\ 
 & Ours & 2.53 $\pm$ 0.20 & 3.24 $\pm$ 0.25 & 2.77 $\pm$ 0.23 & 0.78 \\

\grayrow
& \textsc{Delta} & \textcolor{darkgreen}{\texttt{-}\textbf{45.9\%}} & \textcolor{darkgreen}{\texttt{-}\textbf{42.4\%}} & \textcolor{darkgreen}{\texttt{-}\textbf{47.8\%}} & - \\ \hline

First 25\% (135 frames) & Baseline & 5.31 $\pm$ 0.26 & 7.36 $\pm$ 0.36 & 6.17 $\pm$ 0.33 & - \\ 
\grayrow
 & Ours & 4.49 $\pm$ 0.25 & 6.10 $\pm$ 0.34 & 5.19 $\pm$ 0.31 & 0.78 \\ 

& \textsc{Delta} & \textcolor{darkgreen}{\texttt{-}\textbf{15.4\%}} & \textcolor{darkgreen}{\texttt{-}\textbf{17.1\%}} & \textcolor{darkgreen}{\texttt{-}\textbf{15.9\%}} & - \\ \hline
\grayrow
First 5\% (27 frames) & Baseline & 5.39 $\pm$ 0.27 & 7.39 $\pm$ 0.37 & 6.25 $\pm$ 0.34 & - \\ 
 & Ours & 4.31 $\pm$ 0.27 & 5.70 $\pm$ 0.37 & 4.96 $\pm$ 0.33 & 0.77 \\

\grayrow
& \textsc{Delta} & \textcolor{darkgreen}{\texttt{-}\textbf{20.0\%}} & \textcolor{darkgreen}{\texttt{-}\textbf{22.9\%}} & \textcolor{darkgreen}{\texttt{-}\textbf{20.6\%}} & - \\ \hline

First 1\% (5 frames) & Baseline & 5.33 $\pm$ 0.27 & 7.27 $\pm$ 0.37 & 6.17 $\pm$ 0.34 & - \\ 
\grayrow
 & Ours & 4.54 $\pm$ 0.28 & 5.85 $\pm$ 0.34 & 5.16 $\pm$ 0.33 & 0.76 \\ 

& \textsc{Delta} & \textcolor{darkgreen}{\texttt{-}\textbf{14.8\%}} & \textcolor{darkgreen}{\texttt{-}\textbf{19.6\%}} & \textcolor{darkgreen}{\texttt{-}\textbf{16.4\%}} & - \\ \hline
\grayrow
No Calibration & Baseline & 5.01 $\pm$ 0.17 & 5.75 $\pm$ 0.21 & 5.57 $\pm$ 0.22 & - \\ 

 & Ours & 4.14 $\pm$ 0.21 & 4.93 $\pm$ 0.26 & 4.68 $\pm$ 0.25 & 0.78 \\ 
\grayrow
& \textsc{Delta} & \textcolor{darkgreen}{\texttt{-}\textbf{17.4\%}} & \textcolor{darkgreen}{\texttt{-}\textbf{14.3\%}} & \textcolor{darkgreen}{\texttt{-}\textbf{16.0\%}} & - \\ \hline

\bottomrule[1.5pt]
\end{tabular}
\arrayrulecolor{black}
\end{table}

\textbf{Alpha Parameter Optimization.}
Adjusting the alpha parameter, as shown in Table \ref{Table: Alpha Selection}, reveals that setting alpha to 'auto' significantly enhances model performance. Specifically, the 'auto' setting reduces the MAE by up to 41\% and the RMSE by up to 40\% compared to a fixed alpha of 7.5. With an MAE of $2.53 \pm 0.20$ and an RMSE of $3.24 \pm 0.25$. A fixed alpha fails to account for variability across samples, undermining the model's generalization across diverse user groups. By dynamically aligning the alpha parameter to individual data characteristics, the auto-adjusted setting enhances predictive accuracy and model robustness.

% Table 4: Alpha Selection
\begin{table}[htp]
\centering
\caption{\textbf{Ablation Study on Alpha Selection with TUAT V1} Performance metrics when calibrated by varying alpha on the TUAT V1 dataset with 270 frames calibration, predicted SpO2, and color check.}
\label{Table: Alpha Selection}
\begin{tabular}{ccccc}
\toprule[1.5pt]
\textbf{Alpha Setting} & \textbf{MAE (\%)} & \textbf{RMSE (\%)} & \textbf{MAPE (\%)} & \textbf{Pearson} \\
\hline \hline
\grayrow
$\alpha$=auto & 2.53 $\pm$ 0.20 & 3.24 $\pm$ 0.25 & 2.77 $\pm$ 0.23 & 0.78 \\
$\alpha$=7.5 & 4.28 $\pm$ 0.28 & 5.37 $\pm$ 0.34 & 4.65 $\pm$ 0.30 & 0.77 \\
$\alpha$=2.5 & 3.01 $\pm$ 0.23 & 3.68 $\pm$ 0.27 & 3.34 $\pm$ 0.26 & 0.79 \\
$\alpha$=1 & 3.65 $\pm$ 0.21 & 4.35 $\pm$ 0.25 & 4.12 $\pm$ 0.25 & 0.78 \\
$\alpha$=0.5 & 4.14 $\pm$ 0.21 & 4.93 $\pm$ 0.26 & 4.68 $\pm$ 0.25 & 0.78 \\
\bottomrule[1.5pt]
\end{tabular}
\end{table}

\textbf{Intelligent Sampling Strategy.}
As shown in Table~\ref{Table: Sampling Strategy}, the intelligent sampling strategy, which selects five frames with varying true SpO$_2$ values, significantly improves performance over using the first five sequential frames. Specifically, the MAE decreases from $4.54 \pm 0.28$ to $3.46 \pm 0.22$, representing an improvement of approximately 24\%. Similarly, the RMSE reduces from $5.85 \pm 0.34$ to $4.48 \pm 0.28$, and the MAPE decreases from $5.16 \pm 0.33$ to $3.93 \pm 0.26$. The Pearson correlation coefficient also increases from 0.76 to 0.78. By focusing on frames with greater SpO$_2$ variation, this approach enhances calibration based on physiologically relevant data rather than quantity alone, potentially boosting predictive accuracy and robustness. However, identifying frames with distinct SpO$_2$ levels introduces uncertainty in the data collection process, as the number of frames required is not fixed. This unpredictability may limit the practicality of the strategy in time-sensitive or resource-limited settings, so we did not use it in Table~\ref{Table: Intra Dataset} and Table~\ref{Table: Inter Dataset}.

% Table 5: Intelligent Sampling
\begin{table}[htp]
\centering
\caption{\textbf{Ablation Study on Intelligent Sampling with TUAT V1} Performance metrics when calibrated by five frames on the TUAT V1 dataset with predicted SpO2 and color check.}
\label{Table: Sampling Strategy}
\begin{tabular}{ccccc}
\toprule[1.5pt]
\textbf{Calibration} & \textbf{MAE (\%)} & \textbf{RMSE (\%)} & \textbf{MAPE (\%)} & \textbf{Pearson} \\
\hline \hline
First 5 frames, $\alpha=\text{auto}$ & $4.54 \pm 0.28$ & $5.85 \pm 0.34$ & $5.16 \pm 0.33$ & 0.76 \\
\grayrow
5 frames (different SpO$_2$), $\alpha=\text{auto}$ & $3.46 \pm 0.22$ & $4.48 \pm 0.28$ & $3.93 \pm 0.26$ & 0.78 \\
\bottomrule[1.5pt]
\end{tabular}
\end{table}

\textbf{Training Label Selection.}
According to related work~\cite{nishidate2022rgb}, tissue oxygen saturation (StO$_2$) is strongly correlated with peripheral oxygen saturation (SpO$_2$). As shown in Table~\ref{Table: Training Labels}, using true StO$_2$ labels to fit true SpO$_2$ yields the lowest MAE of $1.98 \pm 0.13$. However, obtaining true StO$_2$ labels is often impractical. When comparing the use of  StO$_2$ and  SpO$_2$ as training labels, the MAE increases slightly from $2.46 \pm 0.25$ to $2.53 \pm 0.20$, an increase of approximately 2.8\%. This minor difference suggests that training directly with SpO$_2$ labels is a viable alternative, enabling end-to-end training without the need for StO$_2$ labels. This approach simplifies the training process while maintaining comparable accuracy.

% Table 6: Different Training Labels + Few Shot Learning
\begin{table}[htp]
\centering
\caption{\textbf{Ablation Study on different training labels with TUAT V1} Performance metrics when trained by varying labels on the TUAT V1 dataset with 270 frames calibration and color check.}
\label{Table: Training Labels}
\begin{tabular}{ccccc}
\toprule[1.5pt]
\textbf{Training Labels} & \textbf{MAE (\%)} & \textbf{RMSE (\%)} & \textbf{MAPE (\%)} & \textbf{Pearson} \\
\hline \hline
\grayrow
True StO2 & 1.98 $\pm$ 0.13 & 2.54 $\pm$ 0.17 & 2.17 $\pm$ 0.15 & 0.89 \\
Predicted StO2 & 2.46 $\pm$ 0.25 & 3.17 $\pm$ 0.30 & 2.72 $\pm$ 0.28 & 0.81 \\
Predicted SpO2 & 2.53 $\pm$ 0.20 & 3.24 $\pm$ 0.25 & 2.77 $\pm$ 0.23 & 0.78 \\
\bottomrule[1.5pt]
\end{tabular}
\end{table}

\textbf{Influence of Incorporating a Color Checker.}
As shown in Table~\ref{Table: ColorCheck}, implementing a color checker notably enhances cross-dataset training and testing performance by improving measurement consistency across diverse conditions. The MAE decreases from $3.60 \pm 0.22\%$ to $3.17 \pm 0.20\%$, a reduction of 11.9\%, while the RMSE improves from $4.57 \pm 0.29\%$ to $4.11 \pm 0.36\%$, reducing by 10.1\%. By calibrating for variations in lighting and camera characteristics, the color checker enables greater generalization across different hardware setups and capture conditions, making it valuable for robust and reliable physiological measurements in varied environments.

% Table 7： Ablation Study on Color Check
\begin{table}[htp]
\centering
\caption{\textbf{Ablation Study on color check with TUAT V1} Performance metrics when trained on THU and TUAT V2 Datasets with or without color check, tested on the TUAT V1 dataset with 270 frames calibration.}
\label{Table: ColorCheck}
\begin{tabular}{ccccc}
\toprule[1.5pt]
\textbf{Color Check} & \textbf{MAE (\%)} & \textbf{RMSE (\%)} & \textbf{MAPE (\%)} & \textbf{Pearson} \\
\hline \hline
\grayrow
 With Color Check & 3.17 $\pm$ 0.20 & 4.11 $\pm$ 0.36 & 3.62 $\pm$ 0.32 & 0.75 \\
 Without Color Check & 3.60$\pm$ 0.22 & 4.57 $\pm$ 0.29 & 4.00 $\pm$ 0.26 & 0.69 \\
\bottomrule[1.5pt]
\end{tabular}
\end{table}

\textbf{Demographic Study.}
Our analysis demonstrates that demographic factors, including skin tone, age, gender, and COVID-19 history—significantly affect the accuracy of camera-based SpO$_2$ measurements. Performance across skin tone groups shows that darker skin tones (SkinTone 5-6) achieve the lowest error rates (MAE = 1.13\%, RMSE = 1.39\%), suggesting greater signal consistency. In contrast, intermediate skin tones (SkinTone 3-4) exhibit slightly higher error rates (MAE = 1.49\%, RMSE = 1.92\%), potentially due to signal variations linked to melanin concentration. Age also plays a role, as older subjects (30+ years) achieve the lowest errors (MAE = 1.15\%, RMSE = 1.46\%), potentially reflecting physiological factors that stabilize photoplethysmographic (PPG) signals in this group. In contrast, younger adults (ages 18-23) show higher error rates (MAE = 1.49\%, RMSE = 1.88\%), possibly due to increased physiological variability.

Gender and COVID-19 history further influence model accuracy. Female subjects achieve slightly better accuracy (MAE = 1.24\%, RMSE = 1.54\%) compared to males (MAE = 1.40\%, RMSE = 1.77\%), suggesting subtle effects of gender-specific physiology on signal quality or model interpretation. Subjects with a history of COVID-19 exhibit higher error rates (MAE = 1.42\%, RMSE = 1.79\%) relative to those without (MAE = 1.25\%, RMSE = 1.58\%), indicating that post-COVID physiological changes may impact SpO$_2$ measurements, likely due to long-term effects on the cardiovascular or respiratory systems.

\begin{table}[htp]
\centering
\caption{Results on THU Dataset by Skin Color, Age, Gender, and COVID-19 Status}
\label{Table: Category}
\begin{tabular}{llcccc} 
\toprule[1.5pt]
Category & Group & \# Video & MAE (\%) & RMSE (\%) & MAPE (\%) \\ 
\midrule
\midrule
\multirow{3}{*}{Skin Color} & SkinTone 1-2 & 30 & 1.22 $\pm$ 0.12 & 1.49 $\pm$ 0.15 & 1.27 $\pm$ 0.13 \\
                            & SkinTone 3-4 & 57 & 1.49 $\pm$ 0.14 & 1.92 $\pm$ 0.17 & 1.57 $\pm$ 0.14 \\
                            & SkinTone 5-6 & 23 & 1.13 $\pm$ 0.20 & 1.39 $\pm$ 0.23 & 1.18 $\pm$ 0.21 \\
\midrule
\multirow{3}{*}{Age} & 18-23 & 49 & 1.49 $\pm$ 0.11 & 1.88 $\pm$ 0.14 & 1.56 $\pm$ 0.12 \\
                     & 24-29 & 36 & 1.27 $\pm$ 0.17 & 1.60 $\pm$ 0.21 & 1.33 $\pm$ 0.17 \\
                     & >=30 & 25 & 1.15 $\pm$ 0.21 & 1.46 $\pm$ 0.24 & 1.20 $\pm$ 0.22 \\
\midrule
\multirow{2}{*}{Gender} & Male & 73 & 1.40 $\pm$ 0.12 & 1.77 $\pm$ 0.14 & 1.47 $\pm$ 0.12 \\
                        & Female & 37 & 1.24 $\pm$ 0.13 & 1.54 $\pm$ 0.16 & 1.29 $\pm$ 0.14 \\
\midrule
\multirow{2}{*}{COVID-19 Status} & Yes & 60 & 1.42 $\pm$ 0.11 & 1.79 $\pm$ 0.13 & 1.49 $\pm$ 0.12 \\
                                 & No & 50 & 1.25 $\pm$ 0.14 & 1.58 $\pm$ 0.18 & 1.31 $\pm$ 0.15 \\
\bottomrule[1.5pt]
\end{tabular}
\end{table}

\section{Discussion}
\label{sec:discussion}

\textbf{Facial Imaging as a Strong Signal for Blood Oxygen Saturation Estimation.}
We have presented results of single-frame estimation of blood oxygen saturation (SpO$_2$) from facial images. A single frame of facial video can provide signal for accurate SpO$_2$ estimation. Unlike prior work~\cite{hoffman2022smartphone}, which relies on both the DC and AC components of the PPG signal, our model leverages forehead skin pixels and color-check data using only one frame at a time (see Fig \ref{fig:NeuralModel}). This approach exhibits robust performance across both intra- and inter-dataset evaluations, achieving an RMSE of 1.69\% on the THU dataset and 3.24\% on the TUAT V1 dataset with minimal calibration frames (Table \ref{Table: Intra Dataset}). Notably, even without calibration, our method reduced MAE by 17.4\% on the TUAT V1 dataset (Table \ref{Table: Calibration Frames}). In cross-dataset evaluations, it recorded a low MAPE of 3.53\% on the TUAT V2 dataset, outperforming some intra-dataset results. Additionally, the Pearson coefficient consistently exceeded 0.75 in various calibration scenarios, underscoring the method’s adaptability to dynamic settings where traditional multi-frame analysis might be less feasible.

% \textbf{Calibration is Necessary.}
% Calibration proves essential in enhancing both the generalizability and performance of our remote blood oxygen measurement system, demonstrating its critical role across diverse experimental conditions.

% \begin{itemize}
%     \item Extensive calibration significantly improves model accuracy, as indicated by a 38.8\% reduction in MAE when using 270 frames on the TUAT V1 dataset. This is further evidenced by the substantial performance boost from incorporating a color check under varying light conditions in the THU dataset, which improved MAE by 25.4\%. These results underscore the necessity of calibration in adjusting to environmental and experimental variabilities.

%     \item The diversity of subject demographics underscores the importance of calibration parameters. When using 270 frames combined with automated alpha adjustments, we achieved the lowest MAE of 2.53\% on the TUAT V1 dataset. Moreover, employing an intelligent sampling strategy that uses as few as 5 frames still yields a notable MAE of 3.46\%, illustrating that even minimal calibration can significantly enhance performance, provided that it is strategically implemented.
% \end{itemize}

\textbf{Calibration is Essential for Generalizability and Performance.} Calibration is crucial for enhancing both the generalizability and performance of our remote blood oxygen measurement system, highlighting its importance across diverse experimental conditions. Extensive calibration significantly boosts model accuracy, reducing MAE by 38.8\% with 270 frames on the TUAT V1 dataset. Similarly, incorporating a color check for calibration under different cameras and varying light conditions in the inter-dataset led to an 11.9\% improvement in performance, emphasizing calibration’s role in adapting to environmental and experimental variability. 

Moreover, the diversity of subject demographics underscores the importance of calibration parameters. For instance, utilizing 270 frames with automated alpha adjustments achieved the lowest MAE of 2.53\% on the TUAT V1 dataset, while an intelligent sampling strategy with only 5 frames still resulted in a notable MAE of 3.46\%, demonstrating that even minimal calibration, when strategically applied, can yield substantial performance gains. These findings collectively affirm that calibration is not only beneficial but necessary for achieving consistent accuracy across dynamic settings.

% \subsection{Neural Methods Yield Better Generalizability and Performance}

% \begin{itemize}
%     \item The baseline method, while sufficient for basic measurements on the THU dataset with an RMSE of 1.71\%, significantly underperforms on the TUAT series where lower SpO2 values are prevalent, with RMSE values exceeding 5\%. This disparity underscores the limitations of traditional methods in handling variations across different datasets, especially in more challenging conditions.
%     \item In contrast, our neural method demonstrates a robust improvement in performance, achieving enhancements ranging from 1.2\% to 42.4\% across almost all datasets and settings. These improvements, documented in Tables \ref{Table: Intra Dataset} and \ref{Table: Inter Dataset}, highlight the superior adaptability and effectiveness of neural approaches in managing diverse and complex measurement scenarios compared to traditional methods.
% \end{itemize}

\textbf{Neural Methods Yield Superior Compaed to Signal Processing Methods.}
The baseline method, while achieving an RMSE of 1.71\% on the THU dataset, falls short in more challenging contexts, particularly in datasets like TUAT, where lower SpO$_2$ values are prevalent, leading to RMSE values above 5\%. This significant performance gap underscores the limitations of traditional approaches in maintaining accuracy across diverse datasets, especially under varying physiological conditions. In contrast, our neural model consistently demonstrates robust adaptability, with performance improvements between 1.2\% and 42.4\% across various datasets and measurement scenarios, as detailed in Tables \ref{Table: Intra Dataset} and \ref{Table: Inter Dataset}. These results affirm the neural method’s superior capacity to generalize effectively across complex, real-world settings where traditional methods often struggle.

% \textbf{Subject Context Matters}
% \begin{itemize}
%     \item Our findings indicate that demographic factors such as skin color, age, and gender have measurable impacts on the accuracy of SpO2 predictions. For instance, females and subjects without COVID-19 records yielded more consistent and accurate results, which could influence the calibration and tuning of models for population-specific applications.
%     \item The variability in performance based on these demographic factors necessitates the development of more sophisticated models that can account for such differences, ensuring fairness and accuracy in health monitoring across diverse populations.
% \end{itemize}

\textbf{Subject Context Matters.}
Our findings indicate that demographic factors, including skin color, age, and gender, have measurable impacts on the accuracy of SpO$_2$ predictions. For instance, female subjects and those without a history of COVID-19 exhibited more consistent and accurate results, suggesting that these factors could inform calibration and tuning strategies for population-specific applications. The observed variability underscores the need for more sophisticated models that can account for demographic differences, ensuring fairness and accuracy in health monitoring across diverse populations.

\textbf{Limitations.}
The dependence on extensive calibration represents a limitation for rapid deployment scenarios, highlighting the need for research into models that require minimal or no upfront calibration. Addressing this issue could enhance the usability and practicality of remote SpO$_2$ monitoring technologies. Although our approach advances non-invasive blood oxygen monitoring, further studies are essential to refine these models to better manage inherent variability in environmental and physiological conditions, especially in less controlled settings beyond laboratory environments.

\section{Conclusion}
Our study demonstrates that deep learning models applied to video data can accurately measure blood oxygen saturation, advancing the potential impact of remote, camera-based health monitoring. Through comprehensive intra- and inter-dataset evaluations, including ablation experiments, our method achieved significant accuracy gains over traditional signal processing approaches, with robust performance across diverse skin tones, age groups, and genders. The findings underscore the importance of calibration for generalizability, especially in dynamic lighting conditions, while showing that even minimal calibration, when intelligently applied, can yield substantial performance improvements. Although the method shows promise, the need for controlled calibration presents a limitation for broader deployment, indicating the value of future research into calibration-free models. In sum, our approach offers a scalable, non-invasive solution for SpO$_2$ monitoring that could support accessible health monitoring in real-world settings.

\section{Methods}
\label{sec:methods}

\subsection{Dataset}
Our study involved the collection of two large datasets of facial videos with carefully synchronized blood oxygen measurements. The datasets were collected using similar protocols, but from different subjects, in different environments and with different hardware. As such we are able to investigate the generalizability of our method.

\begin{figure}[t!]
    \centering
    \includegraphics[width=1\textwidth]{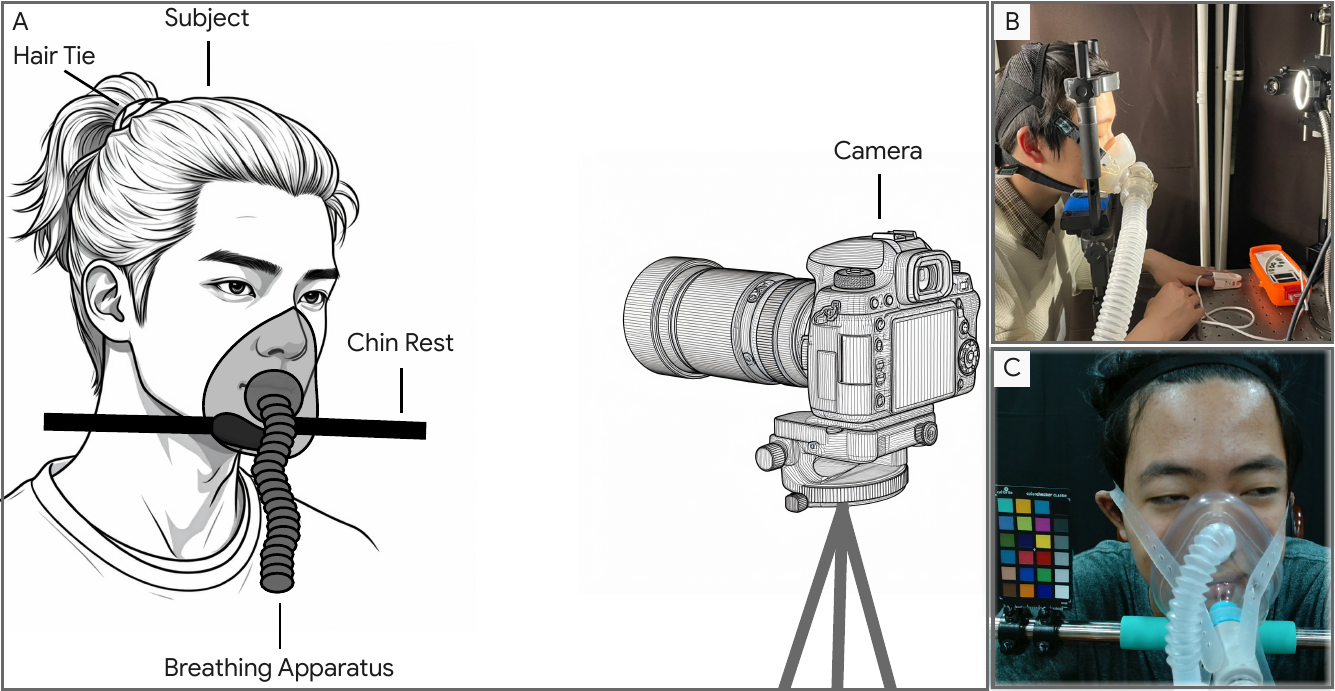} %\\
    \vspace{0.2cm}
    \caption{\textbf{Study Apparatus.} A) Diagram of study apparatus design. B) Example from the TUAT datasets. C) Example from the THU dataset.}
    \label{fig:data_examples}
  \end{figure}

\subsubsection{TUAT Datasets}

\textbf{Imaging System.}
A white light-emitting diode (LED) (LA-HDF158, Hayashi Watch Works Co., Ltd., Tokyo, Japan) illuminated the subjects' facial skin through a ring-shaped illuminator with a light guide (LGC1-8L1000-R55, Hayashi Watch Works Co., Ltd., Tokyo, Japan). The diffusely reflected light is received by a 24-bit RGB charge-coupled device camera (DFK23U618, Imaging source LLC, Charlotte, NC, United States) through an analyzer and a camera lens (FL-CC3516-2M, Ricoh Company, Ltd., Tokyo, Japan) to capture RGB color video at a resolution of 640$\times$480 pixels. The ring polarizing filter mounted on the ring illuminator and analyzer was placed in a crossed Nicols orientation to reduce the specular reflection of light from the skin surface. A standard white diffuser with 99\%  reflectance (SRS-99-020, Labsphere Incorporated, New Hampshire, United States) was used for camera white balance and correction of spatial non-uniformity of illumination. Color video of each subject's face was acquired at a 15 Hz frame rate.  The subject's head was fixed on a head rest (SR-HDR, SR Research Ltd., Ottawa, Canada).

\textbf{Experimental Protocols.}
Fig.~\ref{fig:data_examples} shows the experimental setup used in this study.  Participants were seated for 9 minutes during the experiment. Each subject was exposed to normoxia and hypoxia by inhaling the mixture of O$_2$ and N$_2$ at 11\% of the fraction of inspired oxygen (FiO$_2$) for 4 min using a breathing mask connected to the Douglas bag under spontaneous breathing. The FiO$_2$ value was monitored with an oxygen gas sensor (OM-25MS10; TAIEI DENKI, Tokyo, Japan).  This protocol was approved by the institutional review board of the Tokyo University of Agriculture and Technology (approval numbers 200907-0241 and 210903-0338), and all methods were performed according to the protocol.  Simultaneously with the optical imaging of skin tissue, the time course of SpO$_2$ was measured for each subject using a pulse oximeter (OxyTrue$^{\textregistered}$ A SMARTsat, bluepoint MEDICAL GmbH\&Co.KG; Selmsdorf, Germany) with a finger sensor.

We conducted data collection in two phases using two different setups, referred to as TUAT-V1 and TUAT-V2. In the first phase (TUAT-V1), the working distance from the illuminator to the subject's skin was set at 32 cm, with the camera lens positioned 34 cm away. This configuration was used to collect TUAT-V1, comprising 40 experiments involving 15 healthy volunteers of various ages and both sexes. In the second phase (TUAT-V2), these distances were extended to accommodate full-face video capture, providing a broader view to simulate real-world scenarios. Using this setup, TUAT-V2 was collected, consisting of 12 full-face videos recorded from 5 subjects. This adjustment aimed to capture a wider range of facial responses under similar experimental conditions, enhancing the datasets applicability in more dynamic and practical settings such as daily monitoring.

\begin{table}
\centering
\caption{\textbf{Dataset Summary.} Details of the datasets used in this study.}
\label{Table: Dataset Description}
%\arrayrulecolor[rgb]{0.8,0.8,0.8}
\begin{tabular}{lrrllll} 
\toprule[1.5pt]
Name    & \multicolumn{1}{l}{\# People} & \multicolumn{1}{l}{\# Videos} & Sensor       & Position & Light    & Movement  \\ 
\hline \hline
THU     & 71                               & 110                            & Alpha 7R III & Face     & Flexible & Some Head Motions     \\ 
TUAT v1 & 15                               & 40                             & DFK23U618    & Forehead & Fixed    & Limited Head Motion    \\ 
TUAT v2 & 5                                & 12                             & DFK23U618    & Face     & Fixed    & Limited Head Motion     \\
\bottomrule[1.5pt]
\end{tabular}
\arrayrulecolor{black}
\end{table}

\subsubsection{THU Dataset}

\textbf{Imaging System.}
A DSLR camera (Alpha 7R III, Sony Electronics Inc., Wuxi, China) was used to capture high-resolution images. The camera was equipped with a lens (SEL2470GM, Sony Corporation, Thailand) and operated at 50Hz and 1080P resolution. The subjects were illuminated using a light source (Yuanyue2S, Ra98.5, Opple Lighting Co., Ltd., Jiangsu, China) and an additional ring light (26 cm radius, FSTOP, Nanjing, China) to ensure even lighting conditions. A headset (Type F, Xiaobao, Wuhan, China) was used to stabilize the subject's head at a fixed distance of 40 cm from the camera. The setup was designed to minimize motion artifacts and maintain consistent imaging conditions. A color checker (ColorChecker Classic Mini, Xrite Color Technology Co., LTD, Shanghai, China) was included in the frame to calibrate the color accuracy of the images, ensuring that both the color chart and the subject's face were recorded simultaneously.

\textbf{Experimental Protocols.} Fig.~\ref{fig:data_examples} presents the experimental setup employed in this study, involving 71 healthy volunteers (110 recordings) across sexes, ages (18-41 years), and diverse skin tones. Informed consent and baseline health assessments were obtained from all participants before data collection. During each 9-minute session, participants were seated and wore a pulse oximeter (CMS50E, Contec Medical Systems Co., Ltd, Qinghuangdao, China) on a finger, measuring blood oxygen levels, heart rate, and photoplethysmography (PPG) signals at an approximate sampling rate of 20 Hz. Each participant also wore a custom-designed mask with adjustable gas pressure, with oxygen levels maintained at 21\% to match ambient air. For the initial two minutes, the pressure remained within normal atmospheric levels. From the third minute, it was gradually reduced to 0.6 standard atmospheric pressure, then returned to a standard atmospheric pressure. Lighting and imaging settings were varied across experiments, with detailed parameters recorded to ensure robustness. This protocol received approval from the institutional review board of Tsinghua University (approval number: 20230076), and all procedures conformed to the approved protocol. An emergency button was available at all times for participants to terminate the experiment if discomfort arose.

% \answerTODO - JACK TO COMPLETE DRAFT (SEE IZUMI'S EXAMPLE ABOVE)

% Figure ~\ref{fig:data_examples} shows the experimental setup used in this study. Our experiment includes 71 healthy volunteers (110 recordings) of both sexes of different ages (18-41 years) and all kinds of skin tones. Informed consent and health state were obtained from all participants before data collection. Participants were seated for 9 minutes wearing a pulse oximeter (CMS50E, Contec Medical Systems Co., Ltd, Qinghuangdao, China) on their finger to collect physiological data, including blood oxygen levels, heart rate, and PPG signals, at a sampling rate of approximately 20 Hz during the experiment. The subjects wore a custom-made mask with controllable gas pressure, while the oxygen proportion remained consistent with that of air at 21\%. For the first two minutes, the pressure was kept within a normal range. Starting from the third minute, the pressure was gradually reduced to 0.6 atmospheres and maintained until the sixth minute, after which it was restored to normal for the final three minutes. The light and the shooting module are adjusted in different experiments and the parameters are recorded to validate the robustness of our method. This protocol was approved by the institutional review board of Tsinghua University (approval number: 20230076), and all methods were performed according to the protocol. Additionally, an emergency button was placed on the experimental table, which participants could press at any time to terminate the experiment if they felt uncomfortable.

\subsection{Dataset Processing}

This section outlines the methodologies employed in the preprocessing of video data for the TUAT and THU datasets, with specific attention to frame selection, region of interest segmentation, and normalization\/filtering of oxygen saturation labels.

\textbf{Frame Selection.} To ensure a balance between computational efficiency and the representation of low SpO$_2$ values, a total of 540 frames are selected from each video. For the TUAT dataset, frames are selected at a rate of one frame per second. For the THU dataset, selection begins at the index of the minimal SpO$_2$ value and extends both forward and backward to the maximal SpO$_2$ value. The indices are then resampled to evenly select 540 frames, capturing a broad range of SpO$_2$ fluctuations.
\begin{figure}[t!]
    \centering
    
    \includegraphics[width=0.9\textwidth]{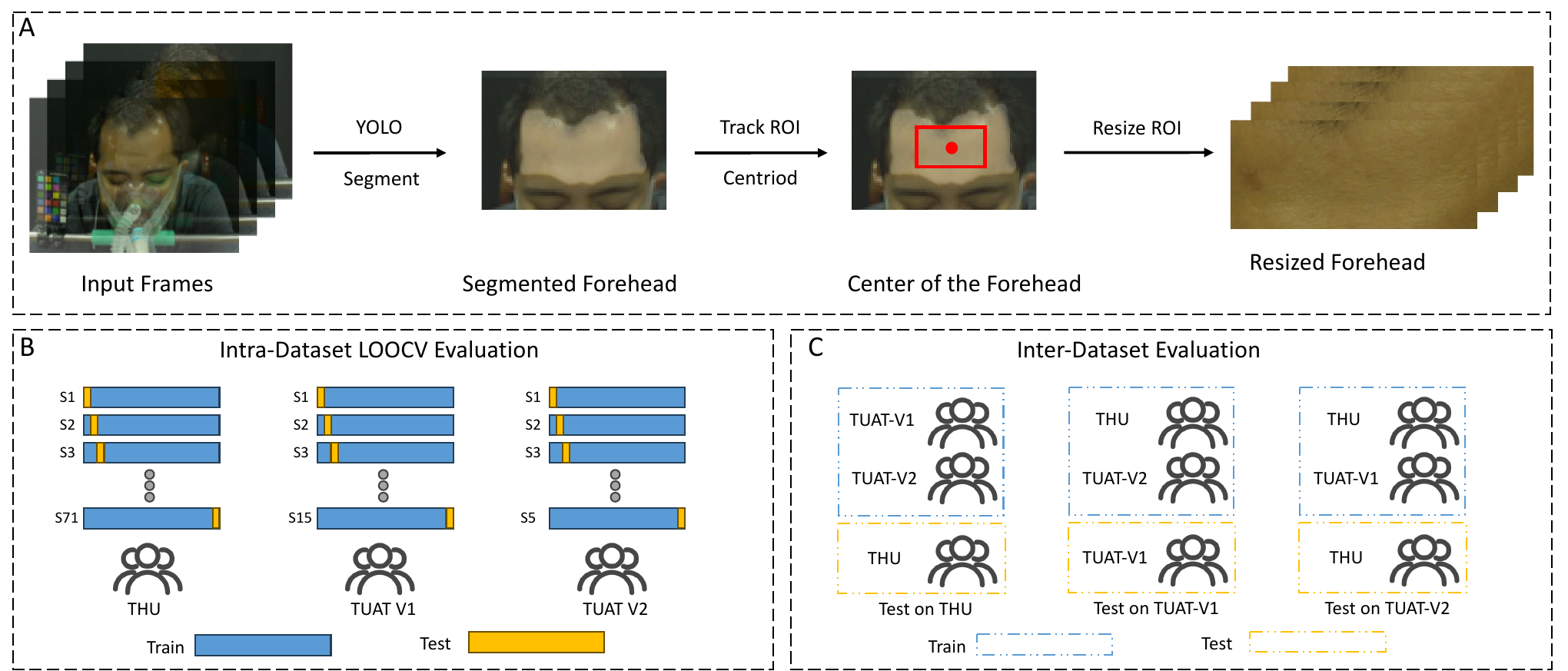}
    \vspace{0.1cm}
    \caption{\textbf{Preprocessing and Evaluation Pipeline.} A) The preprocessing pipeline includes region of interest segmentation, tracking, and resize. B) The Leave-One-Out-Cross-Validation (LOOCV) evaluation pipeline for intra-dataset testing. C) The cross-dataset evaluation pipeline for inter-dataset testing.}
    \label{fig: Preprocess}
\end{figure}

\textbf{Region of Interest Segmentation.} Given that participants wore face masks during the experiments, only the exposed skin areas were analyzed, with the region of interest (ROI) specifically focused on the forehead. As depicted in Fig.~\ref{fig: Preprocess}, we initially manually annotated the forehead regions of 10 sample images from different users to identify key areas. These annotations were then used to train a YOLO-based forehead segmenter via the RoboFlow platform\footnote{\url{https://app.roboflow.com/}}. The trained YOLO model was subsequently applied to automate ROI detection across the remaining images. 

To ensure consistent tracking of the ROI across video frames, particularly during participant movements, the Lucas-Kanade optical flow algorithm was employed~\cite{bruhn2005lucas}. After converting each frame to grayscale, this method tracked the ROI based on the initial frame's segmentation, maintaining spatial consistency. After obtaining the segmented forehead regions, we calculated the geometric centroid of each region and extended outward to form a rectangular ROI. For training and testing within the dataset, the ROI dimensions were set as follows: $300 \times 150$ pixels for the THU dataset, $640 \times 480$ pixels for the TUAT-V1 dataset (no segmentation), and $100 \times 60$ pixels for the TUAT-V2 dataset. For cross-dataset testing, the ROIs of the THU and TUAT-V1 datasets were resized to $100 \times 60$ pixels using OpenCV, ensuring consistency with the TUAT-V2 dataset dimensions.

\textbf{Normalization and Filtering.} Post-processing includes truncating SpO$_2$ values below 80 to 80 and above 100 to 100, ensuring no out-of-range values. Additionally, a low-pass filter with a cutoff frequency of 0.025 Hz is applied to smooth the SpO$_2$ curve, effectively removing unwanted high-frequency noise and enhancing the clarity of physiological signals.

\subsection{Theoretical Basis}

%\subsubsection{Estimation of Tissue Oxygen Saturation}

Supervised learning of deep neural networks involves training a large number of parameters to predict values from high dimensional inputs.  Therefore it is helpful to have a theoretical basis for the potential mapping between inputs (camera frames) and outputs (oxygenation). Each frame of a video recorded by the camera is processed based on a skin tissue model \cite{nishidate2022rgb}. In this model, the \textit{RGB} values in each pixel of the frame are transformed into the Commission Internationale de l’Eclairage (CIE) \textit{XYZ} values with a matrix \textbf{M}$_1$ as
\begin{equation}
    \begin{bmatrix}
    X \\
    Y \\
    Z 
    \end{bmatrix}
    = \textbf{M}_{1}\begin{bmatrix}
    1 \\
    R \\
    G \\
    B 
    \end{bmatrix}
\end{equation}
where
\begin{equation}
\textbf{M}_{1} = \begin{bmatrix}
   \gamma_{0}  &  \gamma_{1} & \gamma_{2}  & \gamma_{3} \\
   \delta_{0}  &  \delta_{1} & \delta_{2}  & \delta_{3} \\
   \epsilon_{0}  &  \epsilon_{1} & \epsilon_{2}  & \epsilon_{3} \\
\end{bmatrix}
\end{equation}
The coefficients \textit{\(\gamma\) }$_i$, \textit{\(\delta\) }$_i$, and \textit{\(\epsilon\) }$_i$, (\textit{i }= 0, 1, 2, 3) are experimentally derived from the measurements of a color checker, which has 24 color standard patches and is provided with data sets containing the \textit{XYZ} values for each patch under specific illuminations and corresponding diffuse reflectance spectra. The concentrations of oxygenated hemoglobin \textit{C}$_{HbO}$, deoxygenated hemoglobin  \textit{C}$_{HbR}$, and melanin  \textit{C}$_{m}$ are then estimated from the \textit{XYZ} values based on the matrix \textbf{M}$_2$. The matrix \textbf{M}$_2$ is generated based on the data set of \textit{XYZ} values derived from the 300 diffuse reflectance spectra in a wavelength range of 400-700 nm at 10 nm intervals, which is numerically derived by Monte Carlo Simulation (MCS) for light transport in skin tissue under various values of  \textit{C}$_{m}$,  \textit{C}$_{HbO}$, and  \textit{C}$_{HbR}$. Multiple regression analysis with 300 data sets established three multiple regression equations as empirical formulas for  \textit{C}$_{m}$,  \textit{C}$_{HbO}$, and  \textit{C}$_{HbR}$:
\begin{equation}
C_{m} =\eta_{0}  +\eta_{1}\textit{X}+\eta_{2}\textit{Y}+\eta_{3}\textit{Z}+\eta_{4}\textit{X}^{2}+\eta_{5}\textit{Y}^{2}+\eta_{6}\textit{Z}^{2}+\eta_{7}\textit{XY}+\eta_{8}\textit{XZ}+\eta_{9}\textit{YZ}
\end{equation}
\begin{equation}
C_{HbO} =\rho_{0}  +\rho_{1}\textit{X}+\rho_{2}\textit{Y}+\rho_{3}\textit{Z}+\rho_{4}\textit{X}^{2}+\rho_{5}\textit{Y}^{2}+\rho_{6}\textit{Z}^{2}+\rho_{7}\textit{XY}+\rho_{8}\textit{XZ}+\rho_{9}\textit{YZ}
\end{equation}
\begin{equation}
C_{HbR} =\sigma_{0}  +\sigma_{1}\textit{X}+\sigma_{2}\textit{Y}+\sigma_{3}\textit{Z}+\sigma_{4}\textit{X}^{2}+\sigma_{5}\textit{Y}^{2}+\sigma_{6}\textit{Z}^{2}+\sigma_{7}\textit{XY}+\sigma_{8}\textit{XZ}+\sigma_{9}\textit{YZ}
\end{equation}
The multiple regression coefficients  \textit{\(\eta\) }$_j$,   \textit{\(\rho\) }$_j$, and   \textit{\(\sigma\) }$_j$, (\textit{j} = 0, 1, 2, ..., 9) here represent the contributions of the \textit{XYZ} values to  \textit{C}$_{m}$,  \textit{C}$_{HbO}$, and  \textit{C}$_{HbR}$, respectively, and are used as the elements of matrix  \textbf{M}$_2$ as
\begin{equation}
\textbf{M}_{2} = \begin{bmatrix}
   \eta_{0}  &  \eta_{1} & \eta_{2}  & \eta_{3} &  \eta_{4} & \eta_{5}  & \eta_{6} &  \eta_{7} & \eta_{8}  & \eta_{9}\\
 \rho_{0}  &  \rho_{1} & \rho_{2}  & \rho_{3} &  \rho_{4} & \rho_{5}  & \rho_{6} &  \rho_{7} & \rho_{8}  & \rho_{9}\\ 
  \sigma_{0}  &  \sigma_{1} & \sigma_{2}  & \sigma_{3} &  \sigma_{4} & \sigma_{5}  & \sigma_{6} &  \sigma_{7} & \sigma_{8}  & \sigma_{9}\\ 
\end{bmatrix}
\end{equation}
The transformation with \textbf{M}$_2$ from the \textit{XYZ} values to \textit{C}$_{m}$,  \textit{C}$_{HbO}$, and  \textit{C}$_{HbR}$ is thus expressed as
\begin{equation}
    \begin{bmatrix}
    C_{m}  \\
    C_{HbO}  \\
    C_{HbR}  
    \end{bmatrix}
    = \textbf{M}_{2}\begin{bmatrix}
    1 \\
    X \\
    Y \\
    Z \\
    X^2 \\
    Y^2 \\
    Z^2 \\
    XY \\
    XZ \\
    YZ \\
    \end{bmatrix}
\end{equation}
Once the matrices \textbf{M}$_1$ and \textbf{M}$_2$ are established, images of  \textit{C}$_{m}$,  \textit{C}$_{HbO}$, and  \textit{C}$_{HbR}$ are reconstructed from an \textit{RGB} image without the MCS. Finally, the tissue oxygen saturation is calculated as
\begin{equation}
StO_{2} \%=100 \times \frac{C_{HbO}}{C_{HbO}+C_{HbR}}
\end{equation}

\subsection{Deep Learning Model}

\subsubsection{Baseline: Estimation of Percutaneous Arterial Oxygen Saturation}

According to historical research~\cite{shapiro2023pulse}, average values of \textit{SpO}\textsubscript{2} have been used as reliable estimates for predicting future \textit{SpO}\textsubscript{2} levels. Building on this foundation, we introduced a user-specific calibration process to achieve a more accurate baseline for each individual. This calibration aims to account for inter-user variability, which can significantly influence prediction accuracy.

To further enhance predictive performance, we optimized the parameters \(\alpha\) and \(\beta\) in a linear regression model for each video. This optimization was performed by minimizing the mean squared error (MSE) between the observed and estimated \textit{SpO}\textsubscript{2} values, ensuring a personalized adjustment to the baseline estimation. By integrating this calibration step, our approach aligns more closely with individual physiological differences, providing a more robust and accurate framework for \textit{SpO}\textsubscript{2} prediction.

We define the model as follows:

\begin{equation}
\alpha, \beta = \arg \min_{\alpha, \beta} \sum_{i=1}^{n} \left( SpO_2^{(i)} - (\alpha \cdot y^{(i)} + \beta) \right)^2
\label{Eqn: calibration}
\end{equation}

In baseline Eqn. \ref{Eqn: calibration}, \(y^{(i)}\) represents the training set blood oxygen saturation values. The parameters \(\alpha\) and \(\beta\) are coefficients that scale and translate the predicted values, respectively, to best fit the actual SpO\(_2\) readings.
The optimal values of \(\alpha\) and \(\beta\) are found through minimization, the model predicts the SpO\(_2\) value for each frame as follows:

\begin{equation}
\widehat{SpO}\textsubscript{2}^{(i)} = \alpha \cdot y^{(i)} + \beta
\end{equation}

\subsubsection{Neural Method}

\begin{figure}[!t]
    \centering
    \includegraphics[width=0.9\linewidth]{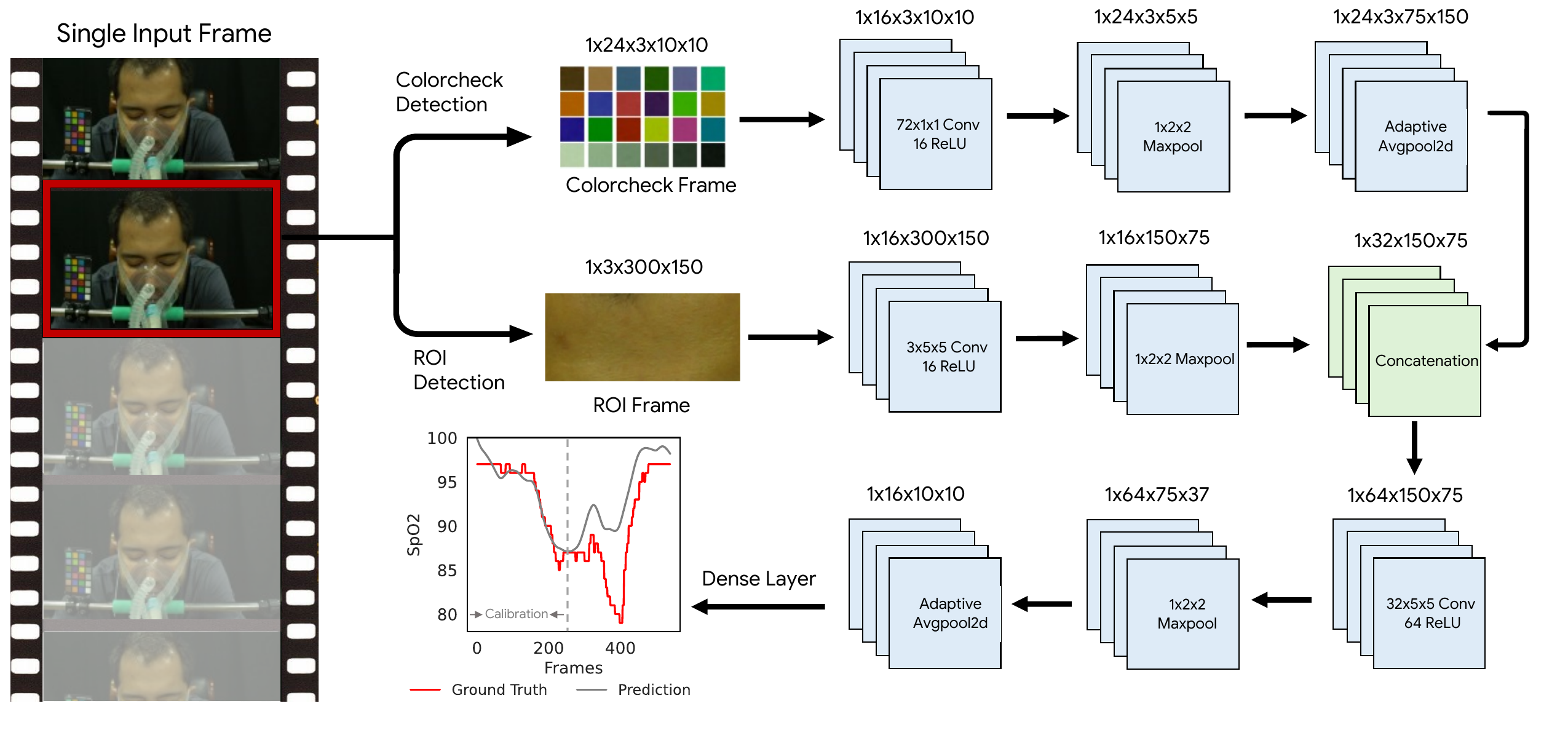}
    \caption{\textbf{VC2S Neural Model.} The model consists of two input branches: a single frame ROI and its corresponding color check. It outputs a SpO$_2$ value for the frame. The model architecture includes 2D convolutional layers with ReLU activation and max pooling, followed by adaptive pooling and a fully connected flattening layer.}
    \label{fig:NeuralModel}
\end{figure}

The VC2S (VideoColorcheckToSpO2, illustrated in Fig.~\ref{fig:NeuralModel}) network integrates video frame inputs and colorcheck data through a dual-path architecture, optimized to extract and fuse complementary features effectively. This design strategy allows the system to process spatial and color information independently, ensuring that each type of data contributes optimally to the final prediction. The video frames are processed using a convolutional layer with 16 filters and a $5\times5$ kernel to capture broad spatial patterns, while the colorcheck data is handled by a $1\times1$ kernel, ideal for isolating precise color details without spatial redundancy.

After the initial convolution, both paths employ ReLU activation functions and $2\times2$ max-pooling layers to reduce the dimensions of the feature maps, thereby retaining only the most salient features and reducing computational load. The colorcheck pathway then undergoes adaptive average pooling to align its dimensions with those of the video pathway, allowing for effective concatenation of features along the channel dimension. This concatenation merges the spatial and color information into a unified representation that is robust to variations in input data and enhances the predictive accuracy of the network.

Further processing is achieved through another convolutional layer with 64 filters and a $5\times5$ kernel, followed by ReLU activation and $2\times2$ max pooling. These layers are designed to refine the combined features, enabling the network to detect complex patterns necessary for accurate \textit{SpO}\textsubscript{2} level predictions. An adaptive average pooling layer reduces the feature map to a manageable size of $10\times10$, setting the stage for the final classification steps. The transition to fully connected layers consolidates the abstracted features, with the first layer reducing the feature dimension to 64 and the final output layer generating a single value indicative of the predicted \textit{SpO}\textsubscript{2} level.

The architectural choices in the VC2S network—including dual-path processing, selective convolution, and strategic pooling—ensure that the model is both efficient and effective in handling different types of data. This approach not only preserves the integrity of each data stream but also optimizes the combination of spatial and color information, resulting in robust performance across varied input scenarios. The network’s design balances complexity with computational efficiency, making it suitable for real-time applications where quick and accurate \textit{SpO}\textsubscript{2} predictions are crucial.

% The VC2S (VideoColorcheckToSpO2, illustrated in Fig \ref{fig:NeuralModel}) network efficiently integrates video frame inputs and colorcheck data through a dedicated dual-path architecture.  Initially, separate convolutional layers extract spatial features from video frames and color information from color checks, allowing each input type to be processed according to its unique characteristics. Following this, ReLU activation and max pooling reduce dimensions and enhance feature representation, while adaptive pooling in the color check path ensures dimensional compatibility with the video features. Once aligned, these features are concatenated, creating a unified representation that captures both textural and color information. Subsequent convolutional, pooling, and adaptive pooling layers further refine this combined feature set, enabling the model to learn complex patterns necessary for accurate SpO2 predictions. This architecture balances the need for complexity and the risk of overfitting, ensuring robust performance across diverse input scenarios while maintaining the integrity of each data type.

\subsubsection{Calibration}
In calibration Eqn.\ref{Eqn: calibration} and Eqn.\ref{Eqn. alpha fit}, \(y^{(i)}\) represents the model prediction of blood oxygen saturation values. \(\alpha\) and \(\beta\) are obtained by minimizing the mean squared error over the first \( n \) frames of actual \(SpO_2\) and predicted \(SpO_2\) values. The final model is:

\begin{equation}
\label{Eqn. alpha fit}
\widehat{SpO_2}^{(i)}= 
\begin{cases} 
\beta + \alpha \times y^{(i)} & \text{if } \alpha > 0 \\ 
\frac{1}{n} \sum_{j=1}^n \text{SpO}_2^{(j)} & \text{if } \alpha \leq 0 
\end{cases}
\end{equation}

After predicting the SpO2 values with our VC2S model, we conducted the auto-calibration process using the start frames. As Eqn. \ref{Eqn. alpha fit} shows that the fitted SpO2 is usually $\beta + \alpha \times y^{(i)}$ unless the alpha is nonpositive, which could happen in some moving and shining frames in the THU dataset.
Performance metrics are computed from the fitted SpO2 values against true values, offering a refined measure of the model's accuracy and reliability in real-world scenarios. This thorough training and post-processing protocol ensures that VC2S not only predicts SpO2 with high fidelity but also adjusts dynamically to the variability inherent in real-world data.

\section{Experiments}

\subsection{Experiment Setup}

All experiments were conducted on a system equipped with an NVIDIA RTX 4090 GPU. The software environment included PyTorch 2.1.1, CUDA 12.1, and Python 3.9.18.

For cross-dataset training, two datasets were used for training, with evaluation on a separate third dataset. In the Leave-One-Out Cross-Validation (LOOCV) setup, the model was trained on videos from all subjects except one, and testing was conducted on the excluded subject's data.

Each model was trained for 15 epochs with a learning rate of \(1 \times 10^{-3}\), using the AdamW optimizer and a Cosine Annealing Learning Rate scheduler. The network's loss function was specifically designed to prioritize accuracy in higher SpO\(_2\) ranges, defined as follows:

\begin{equation}
\text{Loss} = \text{SE}(\text{output} - \text{SpO}_2) \times \left(1 - \frac{\overline{\text{SpO}_2}}{100} \right)
\end{equation}

This custom loss function dynamically adjusts the error contribution based on the average SpO\(_2\) level, emphasizing precision where it is most critical. After training, the predicted SpO\(_2\) values underwent low-pass filtering to smooth out predictions and enhance stability. The model also utilized a fixed alpha value of 2 and calculated beta using the first 10 frames, aligning the model's predictions closely with actual physiological values.

\subsection{Evaluation Details}

\textbf{Cross-Dataset Evaluation.} In the cross-dataset experiments, we evaluated the model on an entirely separate dataset from those used for training. Let \( N \) be the total number of videos in the test dataset. The test dataset is denoted as \( \{ (x_i, y_i) \}_{i=1}^N \), where \( x_i \) represents the features extracted from the \( i \)-th video, and \( y_i \) is the corresponding ground truth SpO\(_2\) value.

\textbf{Leave-One-Out Cross-Validation.} In the LOOCV experiments, we performed leave-one-subject-out validation, where each subject may have one or more videos. In each iteration, all videos belonging to one subject were left out as the test set, and the videos from the remaining subjects were used as the training set. The model was trained on the training set and then made predictions on each video in the test set.

\textbf{Metrics. }For each test video \( i \), the model predicted the SpO\(_2\) value \( \hat{y}_i \) by applying the trained model to the features \( x_i \). The ground truth SpO\(_2\) value is \( y_i \). The errors were computed for each test video \( i \) as follows:

- Mean Absolute Error (MAE):
\[
\text{MAE} = \frac{1}{N} \sum_{i=1}^N | y_i - \hat{y}_i |
\]

- Root Mean Squared Error (RMSE):
\[
\text{RMSE} = \sqrt{ \text{MSE} } = \sqrt{ \frac{1}{N} \sum_{i=1}^N ( y_i - \hat{y}_i )^2 }
\]

- Mean Absolute Percentage Error (MAPE):
\[
\text{MAPE} = \frac{1}{N} \sum_{i=1}^N \left| \frac{ y_i - \hat{y}_i }{ y_i } \right| \times 100\%
\]

- Pearson Correlation Coefficient (PCC):

The Pearson Correlation Coefficient (PCC) is used to quantify the linear relationship between two datasets. Given that oximeter measurements can exhibit a delay of up to 10 seconds, accurate correlation assessment necessitates prior alignment of the datasets.

\[
r = \frac{ \sum_{i=1}^N (y_i - \bar{y})(\hat{y}_{\text{aligned},i} - \bar{\hat{y}}_{\text{aligned}}) }{ \sqrt{\sum_{i=1}^N (y_i - \bar{y})^2} \sqrt{\sum_{i=1}^N (\hat{y}_{\text{aligned},i} - \bar{\hat{y}}_{\text{aligned}})^2} }
\]

where:
\begin{itemize}
    \item \( y_i \) are the data points from the reference dataset (e.g., directly measured physiological data).
    \item \( \hat{y}_{\text{aligned},i} \) are the data points from the oximeter, aligned to compensate for the up to 10-second delay.
    \item \( \bar{y} \) and \( \bar{\hat{y}}_{\text{aligned}} \) are the mean values of the reference and aligned oximeter datasets, respectively:
    \[
    \bar{y} = \frac{1}{N} \sum_{i=1}^N y_i, \quad \bar{\hat{y}}_{\text{aligned}} = \frac{1}{N} \sum_{i=1}^N \hat{y}_{\text{aligned},i}
    \]
\end{itemize}

Aligning the datasets prior to PCC calculation ensures that the temporal discrepancies inherent in oximeter measurements do not skew the correlation results. This method provides a more accurate measure of the true synchronicity and correlation between the datasets. These metrics were aggregated across all test videos to evaluate the overall performance of the model.

% - Pearson Correlation Coefficient (PCC):
% \[
% r = \frac{ \sum_{i=1}^N ( y_i - \bar{y} )( \hat{y}_i - \bar{\hat{y}} ) }{ \sqrt{ \sum_{i=1}^N ( y_i - \bar{y} )^2 } \sqrt{ \sum_{i=1}^N ( \hat{y}_i - \bar{\hat{y}} )^2 } }
% \]
% where:
% \[
% \bar{y} = \frac{1}{N} \sum_{i=1}^N y_i, \quad \bar{\hat{y}} = \frac{1}{N} \sum_{i=1}^N \hat{y}_i
% \]

\newpage
\appendix
\section{Supplement Materials}
\subsection{Table Overview} 
Table \ref{Table: Table Description} provides a comprehensive summary of all tables used in this study, including details on the datasets, calibration parameters, and training labels. These tables offer a structured overview of the data and configurations employed in our experiments, ensuring transparency and reproducibility of the results.

\begin{table}[htp]
\caption{Table Description}
\label{Table: Table Description}
\tiny
\centering
%\arrayrulecolor[rgb]{0.8,0.8,0.8}

\begin{tabular}{lcp{4.8cm}rrrr} 
\toprule[1.5pt]
\multicolumn{7}{c}{Table Explanation}                                                                                                                                                         \\ 

Table & Datasets             & Description                                                                    & Calibration & \# Cal. Frames       & Train Label & Colorcheck    \\ \hline \hline
Table \ref{Table: Intra Dataset}    & 
\makecell{THU \\ TUAT V1 \\ TUAT V2} & Intra-dataset performance with leave-one-subject-out cross-validation (LOSOCV) & Auto               & 270/5                    & SpO2           & w/          \\ 

Table \ref{Table: Inter Dataset}    & \makecell{THU \\ TUAT V1 \\ TUAT V2} & Cross Dataset Performance trained with the other two datasets                  & Auto               & 270/5                    & SpO2           & w/          \\ 
Table \ref{Table: Calibration Frames}    & TUAT V1             & ablation study on calibration frame numbers                                    & Auto               & 0/5/27/135/270             & SpO2           & w/          \\ 
Table \ref{Table: Alpha Selection}    & TUAT V1             & ablation study on calibration alpha                                            & Auto/Fixed         & \multicolumn{1}{r}{270} & SpO2           & w/          \\ 
Table \ref{Table: Sampling Strategy}    & TUAT V1             & ablation study on calibration frame selection strategy                         & Auto               & \multicolumn{1}{r}{5}   & SpO2           & w/          \\ 
Table \ref{Table: Training Labels}   & TUAT V1             & ablation study on Training Labels                                              & Auto               & \multicolumn{1}{r}{270} & SpO2/StO2      & w/         \\ 
Table \ref{Table: ColorCheck}   & TUAT V1                 & ablation study on Colorcheck                                                   & Auto               & \multicolumn{1}{r}{270} & SpO2           & w/+w/o  \\
Table \ref{Table: Category}   & THU                 & demographic study on skin tone, age, gender, and covid                                                   & Auto               & \multicolumn{1}{r}{270} & SpO2           & w/\\
\bottomrule[1.5pt]
\end{tabular}

% \arrayrulecolor{black}
\end{table}

\subsection{Dataset SpO2 Distribution}
\begin{figure}[htp]
    \centering
    \begin{minipage}{0.32\textwidth}
        \centering
        \includegraphics[width=\textwidth]{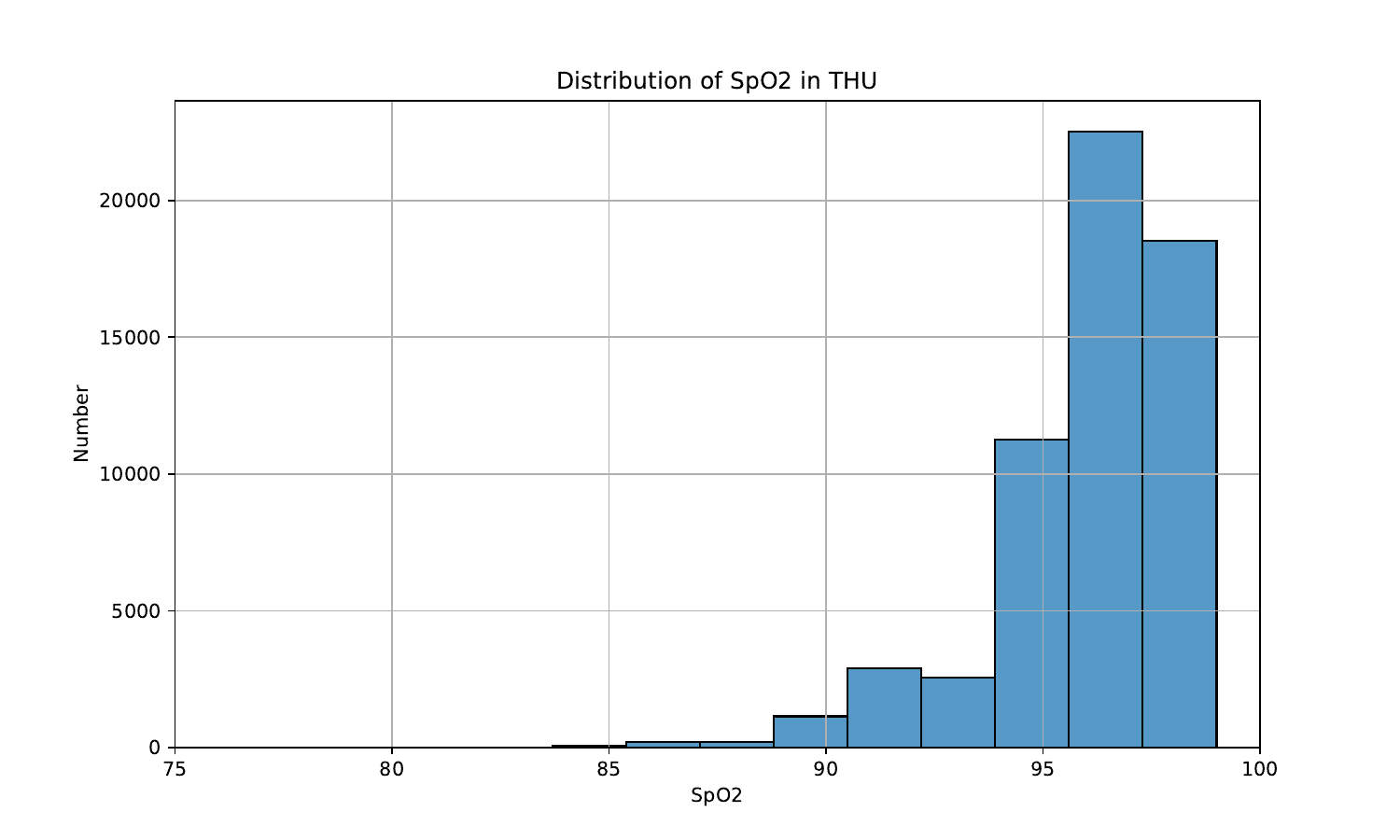}

    \end{minipage}
    \begin{minipage}{0.32\textwidth}
        \centering
        \includegraphics[width=\textwidth]{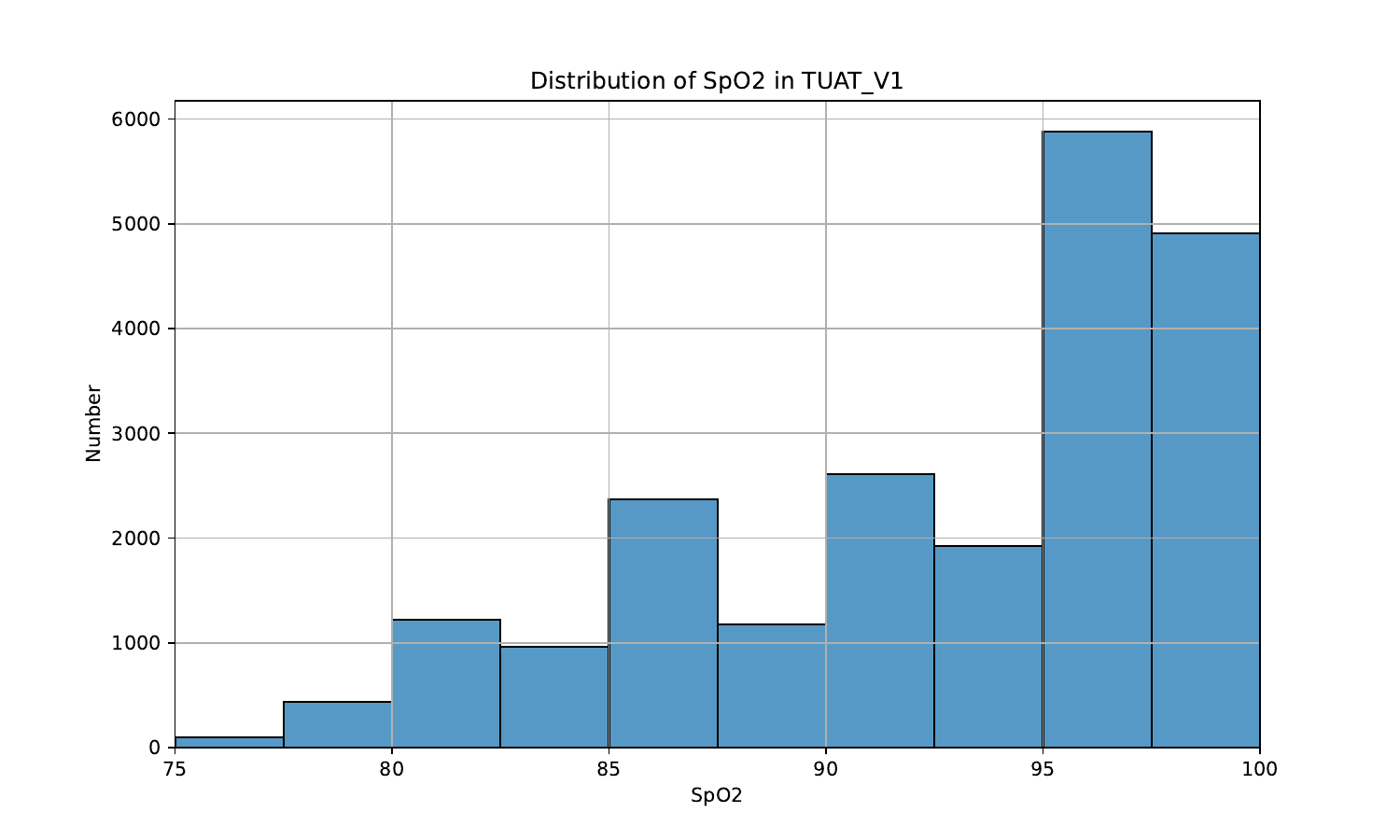}

    \end{minipage}
    \begin{minipage}{0.32\textwidth}
        \centering
        \includegraphics[width=\textwidth]{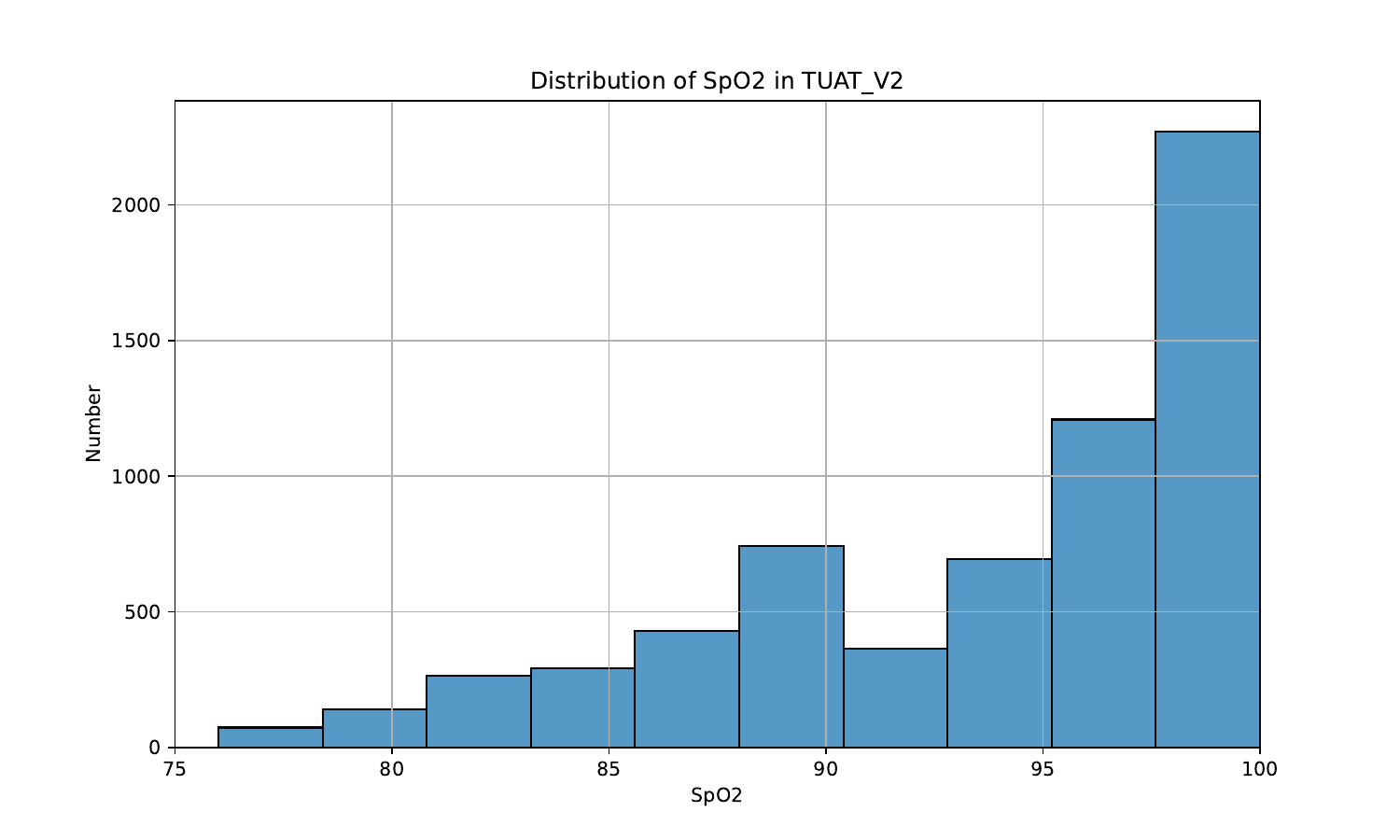}

    \end{minipage}
    \caption{SpO2 distribution in THU, TUAT V1, and TUAT V2 datasets}
    \label{fig:spO2-distribution}
\end{figure}

Figure \ref{fig:spO2-distribution} illustrates the distribution of SpO2 values across the THU, TUAT V1, and TUAT V2 datasets. These visualizations provide insights into the range and variability of SpO2 values in each dataset, highlighting the diversity of physiological conditions captured in the data.

\subsection{Predictions Curves} 
\begin{figure}
    \centering
    \includegraphics[width=1\linewidth]{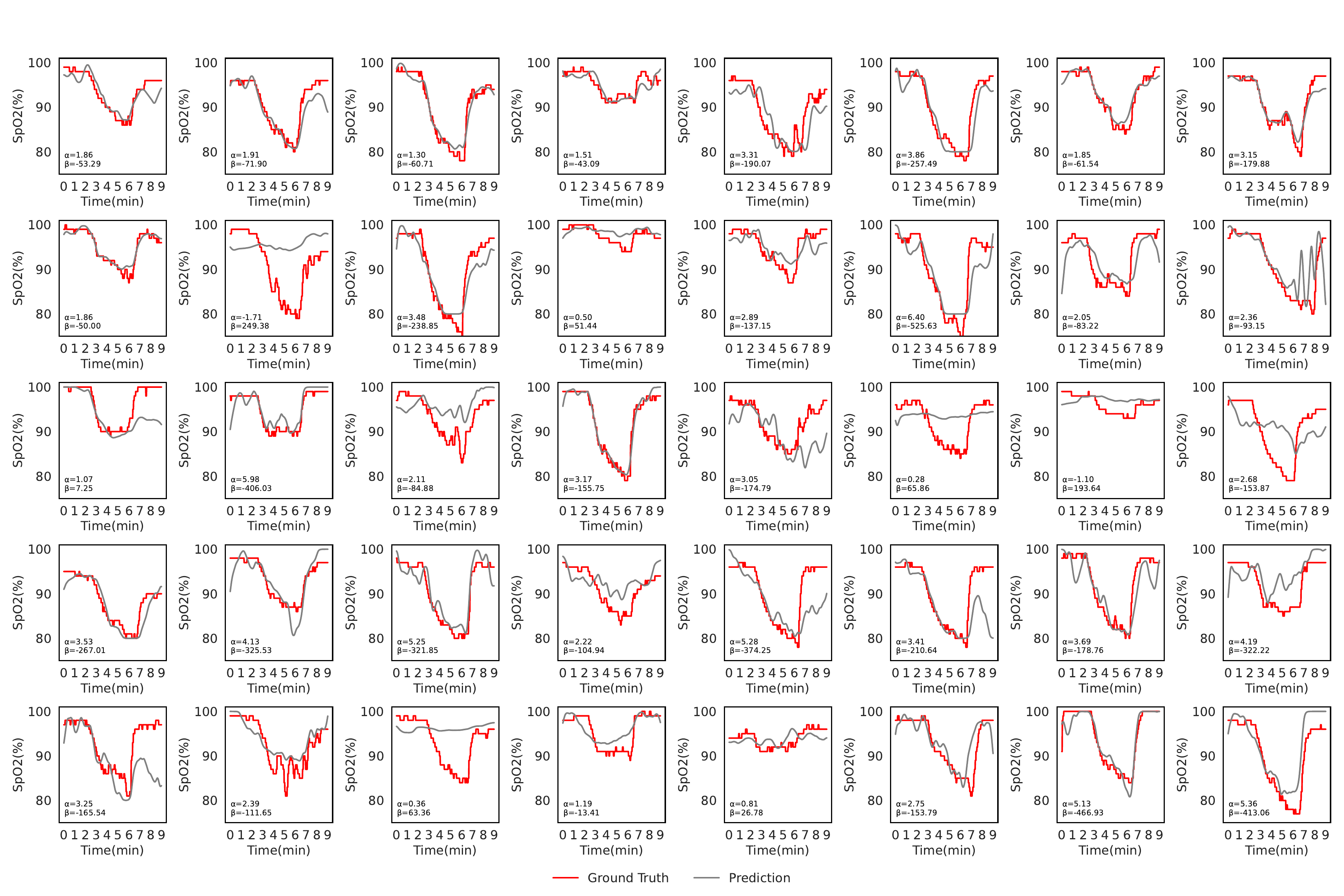}
    \caption{The predictions and labels of the TUAT V1 dataset}
    \label{fig:tuat-v1-all}
\end{figure}

This section presents the predicted and ground truth signals for all videos in the leave-one-out test on the TUAT-V1 dataset. A total of 40 subplots are included in Figure \ref{fig:tuat-v1-all}, illustrating the model's performance across different test cases. These visualizations help in understanding the prediction consistency and accuracy.

\subsection{Code} 
The supplementary materials include the relevant code files and a README document to replicate the main conclusions of this study. Users need to set up the Python environment according to the README instructions and use the datasets referenced in the paper to ensure proper execution and reproducibility of the code.
%\newpage


\bibliography{sn-bibliography}% common bib file


\begin{thebibliography}{28}
\ifx \bisbn   \undefined \def \bisbn  #1{ISBN #1}\fi
\ifx \binits  \undefined \def \binits#1{#1}\fi
\ifx \bauthor  \undefined \def \bauthor#1{#1}\fi
\ifx \batitle  \undefined \def \batitle#1{#1}\fi
\ifx \bjtitle  \undefined \def \bjtitle#1{#1}\fi
\ifx \bvolume  \undefined \def \bvolume#1{\textbf{#1}}\fi
\ifx \byear  \undefined \def \byear#1{#1}\fi
\ifx \bissue  \undefined \def \bissue#1{#1}\fi
\ifx \bfpage  \undefined \def \bfpage#1{#1}\fi
\ifx \blpage  \undefined \def \blpage #1{#1}\fi
\ifx \burl  \undefined \def \burl#1{\textsf{#1}}\fi
\ifx \doiurl  \undefined \def \doiurl#1{\url{https://doi.org/#1}}\fi
\ifx \betal  \undefined \def \betal{\textit{et al.}}\fi
\ifx \binstitute  \undefined \def \binstitute#1{#1}\fi
\ifx \binstitutionaled  \undefined \def \binstitutionaled#1{#1}\fi
\ifx \bctitle  \undefined \def \bctitle#1{#1}\fi
\ifx \beditor  \undefined \def \beditor#1{#1}\fi
\ifx \bpublisher  \undefined \def \bpublisher#1{#1}\fi
\ifx \bbtitle  \undefined \def \bbtitle#1{#1}\fi
\ifx \bedition  \undefined \def \bedition#1{#1}\fi
\ifx \bseriesno  \undefined \def \bseriesno#1{#1}\fi
\ifx \blocation  \undefined \def \blocation#1{#1}\fi
\ifx \bsertitle  \undefined \def \bsertitle#1{#1}\fi
\ifx \bsnm \undefined \def \bsnm#1{#1}\fi
\ifx \bsuffix \undefined \def \bsuffix#1{#1}\fi
\ifx \bparticle \undefined \def \bparticle#1{#1}\fi
\ifx \barticle \undefined \def \barticle#1{#1}\fi
\bibcommenthead
\ifx \bconfdate \undefined \def \bconfdate #1{#1}\fi
\ifx \botherref \undefined \def \botherref #1{#1}\fi
\ifx \url \undefined \def \url#1{\textsf{#1}}\fi
\ifx \bchapter \undefined \def \bchapter#1{#1}\fi
\ifx \bbook \undefined \def \bbook#1{#1}\fi
\ifx \bcomment \undefined \def \bcomment#1{#1}\fi
\ifx \oauthor \undefined \def \oauthor#1{#1}\fi
\ifx \citeauthoryear \undefined \def \citeauthoryear#1{#1}\fi
\ifx \endbibitem  \undefined \def \endbibitem {}\fi
\ifx \bconflocation  \undefined \def \bconflocation#1{#1}\fi
\ifx \arxivurl  \undefined \def \arxivurl#1{\textsf{#1}}\fi
\csname PreBibitemsHook\endcsname

%%% 1
\bibitem[\protect\citeauthoryear{McDuff}{2023}]{mcduff2023camera}
\begin{barticle}
\bauthor{\bsnm{McDuff}, \binits{D.}}:
\batitle{Camera measurement of physiological vital signs}.
\bjtitle{ACM Computing Surveys}
\bvolume{55}(\bissue{9}),
\bfpage{1}--\blpage{40}
(\byear{2023})
\end{barticle}
\endbibitem

%%% 2
\bibitem[\protect\citeauthoryear{Blazek et~al.}{2000}]{blazek2000near}
\begin{bchapter}
\bauthor{\bsnm{Blazek}, \binits{V.}},
\bauthor{\bsnm{Wu}, \binits{T.}},
\bauthor{\bsnm{Hoelscher}, \binits{D.}}:
\bctitle{Near-infrared ccd imaging: Possibilities for noninvasive and
  contactless 2d mapping of dermal venous hemodynamics}.
In: \bbtitle{Optical Diagnostics of Biological Fluids V},
vol. \bseriesno{3923},
pp. \bfpage{2}--\blpage{9}
(\byear{2000}).
\bcomment{International Society for Optics and Photonics}
\end{bchapter}
\endbibitem

%%% 3
\bibitem[\protect\citeauthoryear{Verkruysse
  et~al.}{2008}]{verkruysse2008remote}
\begin{barticle}
\bauthor{\bsnm{Verkruysse}, \binits{W.}},
\bauthor{\bsnm{Svaasand}, \binits{L.O.}},
\bauthor{\bsnm{Nelson}, \binits{J.S.}}:
\batitle{Remote plethysmographic imaging using ambient light.}
\bjtitle{Optics express}
\bvolume{16}(\bissue{26}),
\bfpage{21434}--\blpage{21445}
(\byear{2008})
\end{barticle}
\endbibitem

%%% 4
\bibitem[\protect\citeauthoryear{Takano and Ohta}{2007}]{takano2007heart}
\begin{barticle}
\bauthor{\bsnm{Takano}, \binits{C.}},
\bauthor{\bsnm{Ohta}, \binits{Y.}}:
\batitle{Heart rate measurement based on a time-lapse image}.
\bjtitle{Medical engineering \& physics}
\bvolume{29}(\bissue{8}),
\bfpage{853}--\blpage{857}
(\byear{2007})
\end{barticle}
\endbibitem

%%% 5
\bibitem[\protect\citeauthoryear{Wang et~al.}{2017}]{wang2017algorithmic}
\begin{barticle}
\bauthor{\bsnm{Wang}, \binits{W.}},
\bauthor{\bsnm{Brinker}, \binits{A.C.}},
\bauthor{\bsnm{Stuijk}, \binits{S.}},
\bauthor{\bsnm{Haan}, \binits{G.}}:
\batitle{Algorithmic principles of remote ppg}.
\bjtitle{IEEE Transactions on Biomedical Engineering}
\bvolume{64}(\bissue{7}),
\bfpage{1479}--\blpage{1491}
(\byear{2017})
\end{barticle}
\endbibitem

%%% 6
\bibitem[\protect\citeauthoryear{Chen and McDuff}{2018}]{chen2018deepphys}
\begin{bchapter}
\bauthor{\bsnm{Chen}, \binits{W.}},
\bauthor{\bsnm{McDuff}, \binits{D.}}:
\bctitle{Deepphys: Video-based physiological measurement using convolutional
  attention networks}.
In: \bbtitle{Proceedings of the European Conference on Computer Vision (ECCV)},
pp. \bfpage{349}--\blpage{365}
(\byear{2018})
\end{bchapter}
\endbibitem

%%% 7
\bibitem[\protect\citeauthoryear{Tang et~al.}{2024}]{tang2024camera}
\begin{bchapter}
\bauthor{\bsnm{Tang}, \binits{J.}},
\bauthor{\bsnm{Li}, \binits{X.}},
\bauthor{\bsnm{Liu}, \binits{J.}},
\bauthor{\bsnm{Zhang}, \binits{X.}},
\bauthor{\bsnm{Wang}, \binits{Z.}},
\bauthor{\bsnm{Wang}, \binits{Y.}}:
\bctitle{Camera-based remote physiology sensing for hundreds of subjects across
  skin tones}.
In: \bbtitle{CHI'24 Workshop PhysioCHI'24}
(\byear{2024})
\end{bchapter}
\endbibitem

%%% 8
\bibitem[\protect\citeauthoryear{Castellano~Ontiveros
  et~al.}{2024}]{castellanoontiverosMachineLearningbasedApproach2024}
\begin{barticle}
\bauthor{\bsnm{Castellano~Ontiveros}, \binits{R.}},
\bauthor{\bsnm{Elgendi}, \binits{M.}},
\bauthor{\bsnm{Menon}, \binits{C.}}:
\batitle{A machine learning-based approach for constructing remote
  photoplethysmogram signals from video cameras}.
\bjtitle{Communications Medicine}
\bvolume{4}(\bissue{1}),
\bfpage{109}
(\byear{2024})
\doiurl{10.1038/s43856-024-00519-6}
\end{barticle}
\endbibitem

%%% 9
\bibitem[\protect\citeauthoryear{McDuff}{2018}]{mcduff2018deep}
\begin{bchapter}
\bauthor{\bsnm{McDuff}, \binits{D.}}:
\bctitle{Deep super resolution for recovering physiological information from
  videos}.
In: \bbtitle{Proceedings of the IEEE Conference on Computer Vision and Pattern
  Recognition Workshops},
pp. \bfpage{1367}--\blpage{1374}
(\byear{2018})
\end{bchapter}
\endbibitem

%%% 10
\bibitem[\protect\citeauthoryear{Yu et~al.}{2019}]{yu2019remote}
\begin{bchapter}
\bauthor{\bsnm{Yu}, \binits{Z.}},
\bauthor{\bsnm{Peng}, \binits{W.}},
\bauthor{\bsnm{Li}, \binits{X.}},
\bauthor{\bsnm{Hong}, \binits{X.}},
\bauthor{\bsnm{Zhao}, \binits{G.}}:
\bctitle{Remote heart rate measurement from highly compressed facial videos: an
  end-to-end deep learning solution with video enhancement}.
In: \bbtitle{Proceedings of the IEEE/CVF International Conference on Computer
  Vision},
pp. \bfpage{151}--\blpage{160}
(\byear{2019})
\end{bchapter}
\endbibitem

%%% 11
\bibitem[\protect\citeauthoryear{Liu et~al.}{2020}]{liu2020multi}
\begin{botherref}
\oauthor{\bsnm{Liu}, \binits{X.}},
\oauthor{\bsnm{Fromm}, \binits{J.}},
\oauthor{\bsnm{Patel}, \binits{S.}},
\oauthor{\bsnm{McDuff}, \binits{D.}}:
Multi-task temporal shift attention networks for on-device contactless vitals
  measurement.
NeurIPS
(2020)
\end{botherref}
\endbibitem

%%% 12
\bibitem[\protect\citeauthoryear{Yu et~al.}{2022}]{yu2022physformer}
\begin{bchapter}
\bauthor{\bsnm{Yu}, \binits{Z.}},
\bauthor{\bsnm{Shen}, \binits{Y.}},
\bauthor{\bsnm{Shi}, \binits{J.}},
\bauthor{\bsnm{Zhao}, \binits{H.}},
\bauthor{\bsnm{Torr}, \binits{P.H.}},
\bauthor{\bsnm{Zhao}, \binits{G.}}:
\bctitle{Physformer: Facial video-based physiological measurement with temporal
  difference transformer}.
In: \bbtitle{Proceedings of the IEEE/CVF Conference on Computer Vision and
  Pattern Recognition},
pp. \bfpage{4186}--\blpage{4196}
(\byear{2022})
\end{bchapter}
\endbibitem

%%% 13
\bibitem[\protect\citeauthoryear{Chen and McDuff}{2020}]{chen2020deepmag}
\begin{barticle}
\bauthor{\bsnm{Chen}, \binits{W.}},
\bauthor{\bsnm{McDuff}, \binits{D.}}:
\batitle{Deepmag: Source-specific change magnification using gradient ascent}.
\bjtitle{ACM Transactions on Graphics (TOG)}
\bvolume{40}(\bissue{1}),
\bfpage{1}--\blpage{14}
(\byear{2020})
\end{barticle}
\endbibitem

%%% 14
\bibitem[\protect\citeauthoryear{Nowara et~al.}{2021}]{nowara2021benefit}
\begin{bchapter}
\bauthor{\bsnm{Nowara}, \binits{E.M.}},
\bauthor{\bsnm{McDuff}, \binits{D.}},
\bauthor{\bsnm{Veeraraghavan}, \binits{A.}}:
\bctitle{The benefit of distraction: Denoising camera-based physiological
  measurements using inverse attention}.
In: \bbtitle{Proceedings of the IEEE/CVF International Conference on Computer
  Vision},
pp. \bfpage{4955}--\blpage{4964}
(\byear{2021})
\end{bchapter}
\endbibitem

%%% 15
\bibitem[\protect\citeauthoryear{Tang et~al.}{2023}]{tang2023mmpd}
\begin{bchapter}
\bauthor{\bsnm{Tang}, \binits{J.}},
\bauthor{\bsnm{Chen}, \binits{K.}},
\bauthor{\bsnm{Wang}, \binits{Y.}},
\bauthor{\bsnm{Shi}, \binits{Y.}},
\bauthor{\bsnm{Patel}, \binits{S.}},
\bauthor{\bsnm{McDuff}, \binits{D.}},
\bauthor{\bsnm{Liu}, \binits{X.}}:
\bctitle{Mmpd: multi-domain mobile video physiology dataset}.
In: \bbtitle{2023 45th Annual International Conference of the IEEE Engineering
  in Medicine \& Biology Society (EMBC)},
pp. \bfpage{1}--\blpage{5}
(\byear{2023}).
\bcomment{IEEE}
\end{bchapter}
\endbibitem

%%% 16
\bibitem[\protect\citeauthoryear{Hafen and Sharma}{2018}]{hafen2018oxygen}
\begin{botherref}
\oauthor{\bsnm{Hafen}, \binits{B.B.}},
\oauthor{\bsnm{Sharma}, \binits{S.}}:
Oxygen saturation
(2018)
\end{botherref}
\endbibitem

%%% 17
\bibitem[\protect\citeauthoryear{Nishidate et~al.}{2022}]{nishidate2022rgb}
\begin{barticle}
\bauthor{\bsnm{Nishidate}, \binits{I.}},
\bauthor{\bsnm{Yasui}, \binits{R.}},
\bauthor{\bsnm{Nagao}, \binits{N.}},
\bauthor{\bsnm{Suzuki}, \binits{H.}},
\bauthor{\bsnm{Takara}, \binits{Y.}},
\bauthor{\bsnm{Ohashi}, \binits{K.}},
\bauthor{\bsnm{Ando}, \binits{F.}},
\bauthor{\bsnm{Noro}, \binits{N.}},
\bauthor{\bsnm{Kokubo}, \binits{Y.}}:
\batitle{Rgb camera-based simultaneous measurements of percutaneous arterial
  oxygen saturation, tissue oxygen saturation, pulse rate, and respiratory
  rate}.
\bjtitle{Frontiers in Physiology}
\bvolume{13},
\bfpage{933397}
(\byear{2022})
\end{barticle}
\endbibitem

%%% 18
\bibitem[\protect\citeauthoryear{Shao et~al.}{2015}]{shao2015noncontact}
\begin{barticle}
\bauthor{\bsnm{Shao}, \binits{D.}},
\bauthor{\bsnm{Liu}, \binits{C.}},
\bauthor{\bsnm{Tsow}, \binits{F.}},
\bauthor{\bsnm{Yang}, \binits{Y.}},
\bauthor{\bsnm{Du}, \binits{Z.}},
\bauthor{\bsnm{Iriya}, \binits{R.}},
\bauthor{\bsnm{Yu}, \binits{H.}},
\bauthor{\bsnm{Tao}, \binits{N.}}:
\batitle{Noncontact monitoring of blood oxygen saturation using camera and
  dual-wavelength imaging system}.
\bjtitle{IEEE Transactions on Biomedical Engineering}
\bvolume{63}(\bissue{6}),
\bfpage{1091}--\blpage{1098}
(\byear{2015})
\end{barticle}
\endbibitem

%%% 19
\bibitem[\protect\citeauthoryear{Guazzi et~al.}{2015}]{guazzi2015non}
\begin{barticle}
\bauthor{\bsnm{Guazzi}, \binits{A.R.}},
\bauthor{\bsnm{Villarroel}, \binits{M.}},
\bauthor{\bsnm{Jorge}, \binits{J.}},
\bauthor{\bsnm{Daly}, \binits{J.}},
\bauthor{\bsnm{Frise}, \binits{M.C.}},
\bauthor{\bsnm{Robbins}, \binits{P.A.}},
\bauthor{\bsnm{Tarassenko}, \binits{L.}}:
\batitle{Non-contact measurement of oxygen saturation with an rgb camera}.
\bjtitle{Biomedical optics express}
\bvolume{6}(\bissue{9}),
\bfpage{3320}--\blpage{3338}
(\byear{2015})
\end{barticle}
\endbibitem

%%% 20
\bibitem[\protect\citeauthoryear{Mathew et~al.}{2023}]{mathew2023remote}
\begin{barticle}
\bauthor{\bsnm{Mathew}, \binits{J.}},
\bauthor{\bsnm{Tian}, \binits{X.}},
\bauthor{\bsnm{Wong}, \binits{C.-W.}},
\bauthor{\bsnm{Ho}, \binits{S.}},
\bauthor{\bsnm{Milton}, \binits{D.K.}},
\bauthor{\bsnm{Wu}, \binits{M.}}:
\batitle{Remote blood oxygen estimation from videos using neural networks}.
\bjtitle{IEEE journal of biomedical and health informatics}
\bvolume{27}(\bissue{8}),
\bfpage{3710}--\blpage{3720}
(\byear{2023})
\end{barticle}
\endbibitem

%%% 21
\bibitem[\protect\citeauthoryear{Ding et~al.}{2018}]{ding2018measuring}
\begin{barticle}
\bauthor{\bsnm{Ding}, \binits{X.}},
\bauthor{\bsnm{Nassehi}, \binits{D.}},
\bauthor{\bsnm{Larson}, \binits{E.C.}}:
\batitle{Measuring oxygen saturation with smartphone cameras using
  convolutional neural networks}.
\bjtitle{IEEE journal of biomedical and health informatics}
\bvolume{23}(\bissue{6}),
\bfpage{2603}--\blpage{2610}
(\byear{2018})
\end{barticle}
\endbibitem

%%% 22
\bibitem[\protect\citeauthoryear{Hoffman et~al.}{2022}]{hoffman2022smartphone}
\begin{barticle}
\bauthor{\bsnm{Hoffman}, \binits{J.S.}},
\bauthor{\bsnm{Viswanath}, \binits{V.K.}},
\bauthor{\bsnm{Tian}, \binits{C.}},
\bauthor{\bsnm{Ding}, \binits{X.}},
\bauthor{\bsnm{Thompson}, \binits{M.J.}},
\bauthor{\bsnm{Larson}, \binits{E.C.}},
\bauthor{\bsnm{Patel}, \binits{S.N.}},
\bauthor{\bsnm{Wang}, \binits{E.J.}}:
\batitle{Smartphone camera oximetry in an induced hypoxemia study}.
\bjtitle{NPJ digital medicine}
\bvolume{5}(\bissue{1}),
\bfpage{146}
(\byear{2022})
\end{barticle}
\endbibitem

%%% 23
\bibitem[\protect\citeauthoryear{Liu et~al.}{2024}]{liu2024summit}
\begin{bchapter}
\bauthor{\bsnm{Liu}, \binits{K.}},
\bauthor{\bsnm{Tang}, \binits{J.}},
\bauthor{\bsnm{Jiang}, \binits{Z.}},
\bauthor{\bsnm{Wang}, \binits{Y.}},
\bauthor{\bsnm{Liu}, \binits{X.}},
\bauthor{\bsnm{Li}, \binits{D.}},
\bauthor{\bsnm{Shi}, \binits{Y.}}:
\bctitle{Summit vitals: Multi-camera and multi-signal biosensing at high
  altitudes}.
In: \bbtitle{The 21st IEEE International Conference on Ubiquitous Intelligence
  and Computing (UIC 2024)}
(\byear{2024})
\end{bchapter}
\endbibitem

%%% 24
\bibitem[\protect\citeauthoryear{Wu et~al.}{2023}]{wu2023peripheral}
\begin{barticle}
\bauthor{\bsnm{Wu}, \binits{B.-J.}},
\bauthor{\bsnm{Wu}, \binits{B.-F.}},
\bauthor{\bsnm{Dong}, \binits{Y.-C.}},
\bauthor{\bsnm{Lin}, \binits{H.-C.}},
\bauthor{\bsnm{Li}, \binits{P.-H.}}:
\batitle{Peripheral oxygen saturation measurement using an rgb camera}.
\bjtitle{IEEE Sensors Journal}
\bvolume{23}(\bissue{21}),
\bfpage{26551}--\blpage{26563}
(\byear{2023})
\end{barticle}
\endbibitem

%%% 25
\bibitem[\protect\citeauthoryear{Van~Gastel and
  Verkruysse}{2022}]{van2022contactless}
\begin{barticle}
\bauthor{\bsnm{Van~Gastel}, \binits{M.}},
\bauthor{\bsnm{Verkruysse}, \binits{W.}}:
\batitle{Contactless spo 2 with an rgb camera: experimental proof of calibrated
  spo 2}.
\bjtitle{Biomedical Optics Express}
\bvolume{13}(\bissue{12}),
\bfpage{6791}--\blpage{6802}
(\byear{2022})
\end{barticle}
\endbibitem

%%% 26
\bibitem[\protect\citeauthoryear{Cheng et~al.}{2024}]{cheng2024contactless}
\begin{barticle}
\bauthor{\bsnm{Cheng}, \binits{C.-H.}},
\bauthor{\bsnm{Yuen}, \binits{Z.}},
\bauthor{\bsnm{Chen}, \binits{S.}},
\bauthor{\bsnm{Wong}, \binits{K.-L.}},
\bauthor{\bsnm{Chin}, \binits{J.-W.}},
\bauthor{\bsnm{Chan}, \binits{T.-T.}},
\bauthor{\bsnm{So}, \binits{R.H.}}:
\batitle{Contactless blood oxygen saturation estimation from facial videos
  using deep learning}.
\bjtitle{Bioengineering}
\bvolume{11}(\bissue{3}),
\bfpage{251}
(\byear{2024})
\end{barticle}
\endbibitem

%%% 27
\bibitem[\protect\citeauthoryear{Bruhn et~al.}{2005}]{bruhn2005lucas}
\begin{barticle}
\bauthor{\bsnm{Bruhn}, \binits{A.}},
\bauthor{\bsnm{Weickert}, \binits{J.}},
\bauthor{\bsnm{Schn{\"o}rr}, \binits{C.}}:
\batitle{Lucas/kanade meets horn/schunck: Combining local and global optic flow
  methods}.
\bjtitle{International journal of computer vision}
\bvolume{61},
\bfpage{211}--\blpage{231}
(\byear{2005})
\end{barticle}
\endbibitem

%%% 28
\bibitem[\protect\citeauthoryear{Shapiro et~al.}{2023}]{shapiro2023pulse}
\begin{barticle}
\bauthor{\bsnm{Shapiro}, \binits{I.}},
\bauthor{\bsnm{Stein}, \binits{J.}},
\bauthor{\bsnm{MacRae}, \binits{C.}},
\bauthor{\bsnm{O’Reilly}, \binits{M.}}:
\batitle{Pulse oximetry values from 33,080 participants in the apple heart and
  movement study}.
\bjtitle{NPJ Digital Medicine}
\bvolume{6}(\bissue{1}),
\bfpage{134}
(\byear{2023})
\end{barticle}
\endbibitem

\end{thebibliography}

\end{document}